# High Reversibility of Lattice Oxygen Redox in Na-ion and Li-ion Batteries Quantified by Direct Bulk Probes of both Anionic and Cationic Redox Reactions

Kehua Dai[1, 2], Jinpeng Wu[2, 3], Zengqing Zhuo[2, 4], Qinghao Li[2,5], Shawn Sallis[2, 6], Jing Mao[7], Guo Ai[8], Chihang Sun[1], Zaiyuan Li[1], William E. Gent[9], William C. Chueh[10, 11], Yi-de Chuang[2], Rong Zeng[12], Zhi-xun Shen[3], Feng Pan[4], Shishen Yan[5], Louis F. J. Piper[6], Zahid Hussain[2], Gao Liu[13, *], Wanli Yang[2, 14, *]

[1] School of Metallurgy, Northeastern University, Shenyang 110819, China

[2] Advanced Light Source, Lawrence Berkeley National Laboratory, Berkeley, California 94720, United States

[3] Geballe Laboratory for Advanced Materials, Stanford University, Stanford, California 94305, USA

[4] School of Advanced Materials, Peking University Shenzhen Graduate School, Shenzhen 518055, China

[5] School of Physics, National Key Laboratory of Crystal Materials, Shandong University, Jinan 250100, China

[6] Department of Materials Science and Engineering, Binghamton University, Binghamton, New York 13902, USA

[7] School of Materials Science and Engineering, Zhengzhou University, Zhengzhou 450001, China

[8] College of Physics and Materials Science, Tianjin Normal University, Tianjin 300387, China

[9] Department of Chemistry, Stanford University, Stanford, California 94305, USA

[10] Department of Materials Science & Engineering, Stanford University, Stanford, California 94305, USA.

[11] Stanford Institute for Materials and Energy Sciences, SLAC National Accelerator Laboratory, Menlo Park, California 94025, USA.

[12] Department of Electrical Engineering, Tsinghua University, Beijing 100084, China

[13] Energy Storage and Distributed Resources Division, Energy Technologies Area, Lawrence Berkeley National Laboratory, Berkeley, California 94720, United States

[14] Lead Contact

* Correspondence: wlyang@lbl.gov (W.Y.), gliu@lbl.gov (G.L.)

*1st submission: December 21, 2017*
*2nd resubmission: July 21, 2018*
*Revision: October 21, 2018*

## Summary

The reversibility and cyclability of anionic redox in battery electrodes hold the key to its practical employments. Here, through mapping of resonant inelastic X-ray scattering (mRIXS), we have independently quantified the evolving redox states of both cations and anions in $Na_{2/3}Mg_{1/3}Mn_{2/3}O_2$. The bulk-Mn redox emerges from initial discharge and is quantified by inverse-partial fluorescence yield (iPFY) from Mn-$L$ mRIXS. Bulk and surface Mn activities likely lead to the voltage fade. O-$K$ super-partial fluorescence yield (sPFY) analysis of mRIXS shows 79% lattice oxygen-redox reversibility during initial cycle, with 87% capacity sustained after 100 cycles. In $Li_{1.17}Ni_{0.21}Co_{0.08}Mn_{0.54}O_2$, lattice-oxygen redox is 76% initial-cycle reversible but with only 44% capacity retention after 500 cycles. These results unambiguously show the high reversibility of lattice-oxygen redox in both Li-ion and Na-ion systems. The contrast between $Na_{2/3}Mg_{1/3}Mn_{2/3}O_2$ and $Li_{1.17}Ni_{0.21}Co_{0.08}Mn_{0.54}O_2$ systems suggests the importance of distinguishing lattice-oxygen redox from other oxygen activities for clarifying its intrinsic properties.

## Keywords

## Introduction

The pressing demand of battery technologies with high-capacity, stability and low-cost calls for conceptual breakthroughs and material developments that could be commercially viable. Among the formidable challenges of improving almost every component in batteries, the transition metal (TM) oxide based cathode remains a bottleneck of the energy density and sits at the center of much research and development efforts. In a conventional wisdom, redox reactions associated with only the $3d$ valence states of the $3d$ TMs in oxide cathodes are desirable because involving oxygen in the electrochemical operation was considered detrimental to the reversibility and safety, mainly through the irreversible loss of oxygen and parasitic surface reactions. Such a consideration has directly triggered the innovations of TM oxide based cathodes over chalcogenides, and successfully guided various cathode material developments that are currently employed for commercial batteries.[1] However, recent researches on oxygen states in battery cathodes have indicated that lattice oxygen redox could be reversible and potentially practical for batteries with improved capacity and energy density.[2]

Although the fundamental mechanism of oxygen redox remains elusive and is under active debates between many scenarios, e.g., peroxide with shortened O-O distance,[3,4] a general localized oxygen holes,[5] structural configurations of Alkali-rich and disordered systems,[6] and specific oxygen chemical bonds,[7] a number of claims of oxygen redox have been made in both Li-ion batteries (LIBs) and Na-ion batteries (NIBs) (See a review by Assat and Tarascon,[8] and many references therein).

While the concept of "reversible" oxygen redox has been employed to many cathode materials, as summarized in multiple reviews,[7-9] one could easily notice a mismatch between the claimed "reversibility" and experimental findings. Majority of these previous studies are based on structural probes and hard X-ray absorption of TM $K$-edges, as indirect evidences of the oxygen chemistry. The popular O-$K$ spectroscopy, especially X-ray photoelectron spectroscopy (XPS)[4,10,11] with probe depth of up to 30 nm with hard X-rays,[12] and X-ray absorption spectroscopy (XAS),[5,13-16] have been mostly reported for initial cycles, with only a few studies extended to the second cycle.[17,18] Quantifications of oxygen redox during the first cycle has been performed based on the intensity of the pre-edge features of O-$K$ XAS;[5,13,19] however, the results display obvious discrepancy with the electrochemical profile. For example, the oxygen-redox quantification values display a significant increase even before the charge plateau of the Li-rich compounds.[5,13] Additionally, compared with the initial charge process, the lower-capacity discharge process shows an even higher oxygen-reduction value.[5,13,19] Such discrepancy is due to the fact that intensities of O-K XAS pre-edge features are dominated by the TM-O hybridization, which enhances upon charging due to the oxidization of the system; so it is rather a probe of TM-O hybridization, and does not follow the electrochemical profile on oxygen redox itself.[20] As a matter of fact, the missing reliable experimental results for quantitatively assessing the reversibility of oxygen redox have led to controversy debates. For example, manganese-based NIB cathode materials with low-valence dopants, Na$_y$M$_x$Mn$_{1-x}$O$_2$ ($x$, $y \leq 1$, M = Ni, Mg, Li, etc.), have been reported with high volumetric energy density.[21-24] While reversible oxygen redox has been claimed in several works without direct evidence over extended cycles, other reports suggest that it is challenging to realize reversible oxygen redox with 3$d$-TM based NIB materials.[15,25-27] Furthermore, oxygen gas release through mass spectrometry detects only the irreversible oxygen evolution,[5,13,19,28] and the true relationship between oxygen release and lattice oxygen redox remains unclarified.[28,29] Therefore, a direct and reliable quantification of the oxygen chemistry in the bulk lattice over extended electrochemical cycles is critical for truly assessing the reversibility of lattice oxygen redox in TM oxide cathodes.

Technically, a direct and reliable detection of the intrinsic oxygen redox is a non-trivial issue. A quantitative probe of the lattice oxygen with truly elemental (oxygen) and chemical (state) sensitivity is required. While popular soft X-ray O-*K* studies of XPS and XAS have inspired important debates on oxygen redox,[2-6,13] they suffer technical limitations from their probe depths (6.3 nm with 1.487 keV, 29 nm with 6.9 keV source for XPS[12]) and overwhelming effect from TM-O hybridization in XAS, as briefly explained above.[20] With such concerns and the need of a reliable characterization of lattice oxygen redox, we have recently developed ultra-high efficiency soft X-ray full-range mapping of resonant inelastic X-ray scattering (mRIXS),[30] which was soon demonstrated to be a sensitive and a reliable probe of the TM and oxygen redox in bulk battery electrodes beyond conventional XAS.[31-33] Compared with XAS, mRIXS further resolves the energy of the fluorescence photon, called emission energy, which is completely missing in XAS.[20] In particular, a specific O-*K* mRIXS feature at 523.7 eV emission energy, different from that of TM-O hybridization feature at 525 eV, emerges only from the high-voltage charge plateau in Li-rich compounds,[31] and only in oxygen redox systems.[33] This feature is buried in conventional XAS lineshape,[31,33] but manifests itself in the scanning transmission X-ray microscopy (STXM) only when the probe is moved away from the surface for tens of nm.[31] It is important to note that although the probe depth is the same (100 nm) for both XAS in fluorescence mode and for mRIXS, mRIXS does not suffer the lineshape distortion as in fluorescence XAS, and the bulk sensitivity of mRIXS for probing oxygen redox is further enhanced by resolving emission energies, i.e., bulk and surface signals fortunately display different emission energies in mRIXS. For example, comparison with STXM of Li-rich materials directly show that signals corresponding to the mRIXS oxygen-redox feature only show up when the probe is moved deeper than tens of nm, i.e., the 523.7 eV emission energy mRIXS signal fingerprints lattice oxygen redox not only due to the probe depth, but also through the specific emission energy values.[31] While STXM provide important proof of the bulk sensitivity of mRIXS through emission energies, it requires TEM-type of nanoparticle samples and mRIXS provides the important advantages on being able to measure all electrodes in any forms.

Furthermore, recent combined studies of mRIXS experiments and theoretical calculations based on OCEAN package of $Li_2O_2$ show that the mRIXS feature at 523.7 eV emission energy stems from a specific O *2p-2p* intra-band exciton.[34] Such excitations require a partially unoccupied O-*2p* orbital (so electrons could be excited to the unoccupied *2p* states), which naturally fingerprints the oxidized oxygen states because $O^{2-}$ has no unoccupied *2p* state. Although the specific excitations in the more complex transition-metal systems are yet to be revealed with further works, these experimental and theoretical studies demonstrate that the specific mRIXS feature is a reliable fingerprint of the oxidized oxygen state with enough elemental and chemical sensitivity to differentiate the lattice oxygen redox signals in battery electrodes from the dominating hybridization features.[20] It is also important to note that mRIXS images reveal the full profile of oxygen redox features along both the excitation and emission energies. Such a full profile is particularly important for a reliable quantification because different systems of oxidized oxygen may display different profile, e.g., $Li_2O_2$ shows a long profile along excitation energies, which cannot be detected through individual cuts in conventional RIXS experiments.[5,13,19] Therefore, these mRIXS developments and observations have now set a solid foundation for quantitative analysis of the lattice oxygen redox in batteries.

On the other hand, almost all the oxygen-redox active electrodes involve cationic redox reactions, simultaneously or separately.[35,36] As directly shown later in this work, even for a system with nominally pure oxygen redox, TM redox will emerge with electrochemical cycling and affects the oxygen redox. Direct probes of TM redox, together with oxygen redox, will provide a consistent and experimental clarification of the reaction mechanism responsible for the total capacity, as well as the critical relationship between the anionic and cationic redox. The TM redox has more often been measured by hard X-ray XAS of the $K$-edges, through the edge shifting of the $2s$-$4p$ excitation features.[37] However, for $3d$ TMs, soft X-ray TM-$L$ edges correspond to the excitations from $2p$ to directly the valence $3d$ states, providing a more direct probe of the TM-$3d$ states,[37,38] which is more sensitive to valence electron states and could be easily quantified.[39,40] Unfortunately, the bulk-probe mode, namely fluorescence yield (FY), of the TM-$L$ XAS, is known to be distorted due to the self-absorption effect,[41] leading to complicated situations that require theoretical calculations for quantitative analysis.[42] In 2011, Achkar et al. has demonstrated soft X-ray inverse partial fluorescence yield (iPFY) as a non-distorted bulk probe of TM-$3d$ states.[41] Such a technique has now been much improved through the 0.2~0.4 eV energy resolution of mRIXS, which enables the iPFY technique for all TM elements involved in practical battery cathodes, especially Mn.[30] Obtaining non-distorted bulk TM-$L$ spectra through mRIXS opens up the opportunity of a straightforward quantification of the bulk TM redox based on directly the TM-$3d$ state measurements.

In this work, we further developed mRIXS technique and demonstrate direct and reliable quantifications of the evolution of both the oxygen and TM redox states upon hundreds of cycles in two contrasting electrode systems. As shown in Fig. 1 and elaborated later, by extracting the super-partial fluorescence yield (sPFY) within the characteristic emission-energy range of oxygen redox in O-$K$ mRIXS, the variation of lattice oxygen redox states could be quantified at different stages and at different cycles of electrochemical cycles. In the meantime, the extracted iPFY data from TM-$L$ mRIXS allow quantitative definition of the bulk TM oxidation states based on established linear-fitting method.[39] The combination of the final analysis of O-$K$ mRIXS-sPFY and Mn-*L* mRIXS-iPFY provides a complete picture of the evolution of both the anionic and cationic redox in $Na_{2/3}Mg_{1/3}Mn_{2/3}O_2$ at different potentials and for a hundred cycles.

The material we focused on in this study is $Na_{2/3}Mg_{1/3}Mn_{2/3}O_2$, with its stoichiometry deliberately adjusted based on the $Na_{2/3}[Mg_{0.28}Mn_{0.72}]O_2$ electrodes previously reported by Yabuuchi et al.[43] Our adjusted stoichiometry increases the nominal valence of Mn to 4+, i.e., $1_{Na}\times2/3+2_{Mg}\times1/3+4_{Mn}\times2/3=2_O\times2$ to minimize the Mn redox during the initial charge process. Compared with the previous report,[43] our material displays an even higher initial charge capacity ($162$ mAh g$^{-1}$) and a much larger contribution from the high-voltage charge plateau. We note that because cathode is the source of Na-ions in common NIBs, the initial charge capacity of the cathode is practically important for NIB full cells. Considering the high valence Mn in the pristine material and the high initial charge capacity, this re-designed stoichiometry of $Na_{2/3}Mg_{1/3}Mn_{2/3}O_2$ makes the material a perfect candidate to study the evolution (reversibility) of oxygen redox. Additionally, we employed the mRIXS-sPFY analysis to a typical Li-rich compound, $Li_{1.17}Ni_{0.21}Co_{0.08}Mn_{0.54}O_2$, as reported previously[31], which successfully reveals the reversivility and cyclability of the lattice oxygen redox activities in Li-rich electrode with known gas release and surface reactions involved.

Our final results of both the oxygen and TM redox quantifications are self-consistent and deciphers the total capacity of $Na_{2/3}Mg_{1/3}Mn_{2/3}O_2$. The data show that oxygen redox is 79% reversible during the initial charge/discharge cycle, and remains stable up to 50 cycles. After 50 cycles, 87% capacity is sustained after a hundred cycles. The quantifications also show that cationic (Mn) redox in $Na_{2/3}Mg_{1/3}Mn_{2/3}O_2$ emerges during the first discharge and develops low-valence $Mn^{2+}$ on surface. Recently observations show that TM valence states continuously drops upon cycling redox in Li-rich electrodes, contributing to voltage fade.[44] Here, evolution towards low valence Mn in bulk is small, but the increase of low-valence Mn, especially $Mn^{2+}$, is clearly observed on the electrode surface, which is largely responsible for the voltage fade. Furthermore, the same mRIXS-sPFY analysis reveals a surprisingly high initial-cycle reversibility of the lattice (bulk) oxygen redox reaction in $Li_{1.17}Ni_{0.21}Co_{0.08}Mn_{0.54}O_2$ (76%), considering the system is known to involve serious gas release and surface reactions, but with only 44% capacity retention after 500 cycles. Our results demonstrate direct (Mn-3*d*, O-2*p*) and reliable experimental quantification of the changing cationic and anionic redox states in bulk lattice, and indicate that NIB and LIB materials share the same fundamental mechanism of oxygen redox. The lattice oxygen redox displays high reversibility but capacity decay is obvious in systems involving other non-lattice oxygen activities. The contrast between the two different systems indicates that our conventional understanding of oxygen redox was confuses by mixing lattice oxygen redox with oxygen release and/or surface reactions. While other redox active oxygen activities is irreversible, lattice oxygen redox could be highly reversible for hundreds of cycles in both Na-ion and Li-ion batteries.

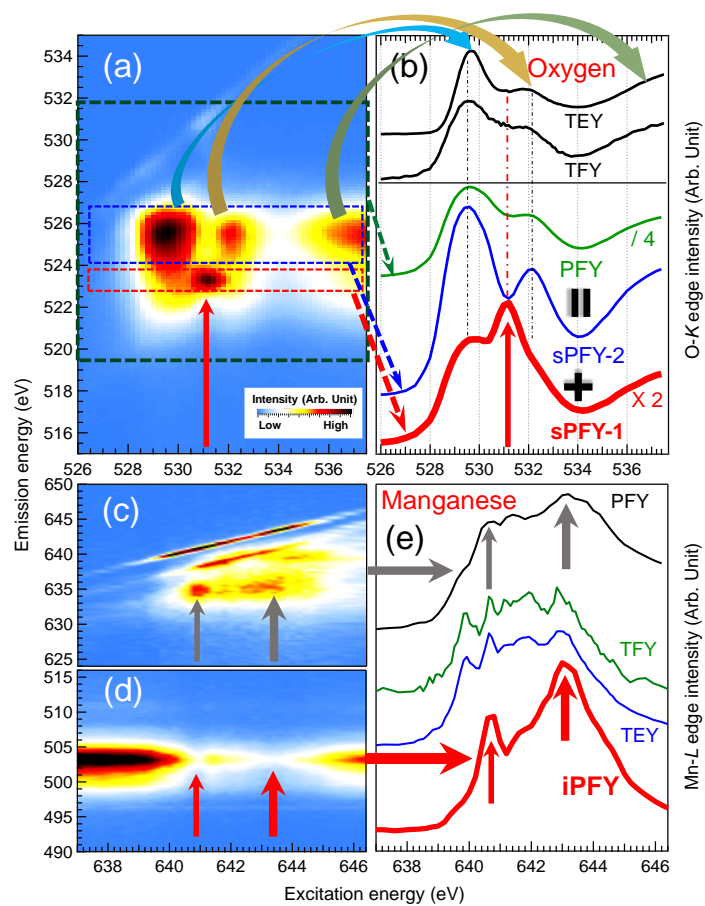

**Fig. 1 | mRIXS of O-*K*/Mn-*L* and their sPFY/iPFY analysis for a direct probe of bulk redox.** (a) O-*K* mRIXS of a typical $Na_{2/3}Mg_{1/3}Mn_{2/3}O_2$ NIB electrode at charged state. sPFY spectra (red) plotted in (b) are extracted by vertically integrating the signal across a super-partial range marked by the red (sPFY-1) and blue (sPFY-2) frames in (a). Conventional O-*K* PFY (green frame), TFY and TEY data of XAS are plotted in (b) for comparison. mRIXS-sPFY selectively integrates the mRIXS features in a specific emission-energy region, so the sPFY peak area provide the opportunity for quantifying the interested mRIXS features in both excitation and emission energy dimensions, something cannot be achieved through individual cuts in conventional RIXS. (c) and (d) are mRIXS images with excitation energy at the same Mn-*L* edge range, but with emission energy at (c) Mn-*L* and (d) O-*K* edges. The diagonal features in the top left part of (a) and (c) are elastic features used for energy calibration. Integration of the intensity in (c) leads to the Mn-*L* PFY spectrum in (e). The inverse of the integrated intensity of (d) is the Mn-*L* iPFY, which is a non-distorted FY bulk probe of the Mn-3*d* valence states. Conventional TFY and TEY data of Mn-*L* XAS are plotted in (e) for comparison.

## Results and discussion

### mRIXS, sPFY of O-*K*, and iPFY of Mn-*L*

For a better readability and technical validity, we first explain our technical approaches for direct probes of the O-2$p$ and TM-3$d$ valence states through O-*K* and Mn-*L* mRIXS, respectively. We also explain the sPFY and iPFY analysis of the mRIXS results for the quantifications of the oxygen and TM states. Although iPFY itself and its advantages over conventional XAS has been introduced before,[41] iPFY through mRIXS provides much more information on the relevant spectroscopic excitations due to its greatly improved resolution compared with previous reports.[20] It is important to note that, although sPFY is plotted upon the same excitation axis of XAS for comparison purpose, features of sPFY is fundamentally different from XAS features as they are associated with a very narrow (super-partial) emission energy range. i.e., mRIXS-sPFY is plotted in the way of XAS for readability purpose, but sPFY features are of RIXS character, not XAS.

Fig. 1a shows a typical O-*K* mRIXS result collected on a charged electrode (sample detailed below), and its sPFY analysis in Fig. 1b. mRIXS shows the intensity (in color scale) distribution upon both the excitation (horizontal axis, same as the energies in XAS) and emission energies (vertical axis, missing in XAS).[20] The sum of all the fluorescence counts along vertical axis is what conventional XAS collects in total fluorescence yield (TFY) mode. Integrating within the main O-*K* signal range (519-529 eV emission energy) gives the typical partial fluorescence yield (PFY). Consistent with all previous publications, the XAS in PFY, TFY and TEY (total electron yield) (Fig. 1b) consists of 3 main features corresponding to the 3 intensity packets of fluorescence signals in mRIXS around 525 eV emission energy (top arrows in Fig. 1). Although large energy scale mRIXS has only been realized recently,[30] these three broad features along 525 eV emission energy have been identified as features from the O-2$p$ bands (>534 eV excitation energy) and its hybridizations to TMs (529 and 532 eV excitation energies) through conventional RIXS cuts, as previously reviewed in fundamental physics.[45] Such an assignment could also be directly seen in Fig. 1, because their excitation energies match exactly the conventional XAS features with the same assignment, i.e., features in conventional XAS are dominated by only the 525 eV emission-energy features from standard oxygen (O$^{2-}$) signals.

Consistent with the oxygen-redox feature identified in LIB systems,[20,31,33] mRIXS clearly shows that a sharp feature at 523.7 eV emission energy (red arrows in Fig. 1). RIXS features at 523.7 emission energy has been found in oxidized oxygen systems, e.g., O$_2$ gas and peroxides.[34,46] The excitation energy of this feature, 531 eV, is within the energy range of those of non-divalent oxygen species, however is much sharper compared with peroxides and is material dependent.[34] Therefore, this O-*K* mRIXS feature at 523.7 eV emission energy and 531 eV excitation energy provides a reliable spectroscopic fingerprint of the non-divalent oxygen, i.e., oxidized oxygen, in battery electrodes at charged states.

By integrating the intensity across the narrow emission-energy range around the characteristic emission energies of 523.7 eV (Red frame in Fig. 1a) and 525 eV (Blue frame in Fig. 1a), a super-partial fluorescence yield (sPFY) signals quantify the intensity of the specific mRIXS features from specific origins. As shown in the bottom of Fig. 1b, the sPFY-1 of the 523.7 eV emission energy range (red) displays a distinct peak at 531 eV excitation energy, which represents the oxygen redox activities. However, all conventional XAS lineshapes are dominated by the broad and strong features around 525 eV emission energy, as shown by sPFY-2 (blue frame). As explained above, RIXS features at 525 eV emission are of standard $O^{2-}$ character from both broad O-2*p* bands and TM-O hybridization.[45] Therefore, O-*K* mRIXS successfully distinguishes the lattice oxygen redox signals from other strong contributions. More importantly, through a full energy range map along both emission energy and excitation energies in mRIXS, sPFY around 523.7 eV emission energy (Red frame in Fig. 1a) contains the full-range mRIXS intensity along emission energy, which enables a quantification of the total oxygen-redox signals if the sPFY is further integrated along excitation energies, as demonstrated below.

A technical note is that X-ray radiation may affect the oxidized oxygen species,[47] which may decrease the oxygen-redox signals. Careful radiation dose tests found that $Na_{2/3}Mg_{1/3}Mn_{2/3}O_2$ is stable across a large X-ray dose range, but Li-rich compounds display some decrease of the oxygen-redox feature intensity upon high X-ray dose, which will be reported separately. Although we have carefully controlled the radiation dose with a relatively relaxed beam size of 20-30 μm for mRIXS experiments, there is still a possibility that the oxygen-redox intensity of $Li_{1.17}Ni_{0.21}Co_{0.08}Mn_{0.54}O_2$ is underestimated in experiments due to irradiation effect. However, as elaborated later, the majority of the contrast between the two systems are undoubtedly dominated by the intrinsic differences of the redox-active oxygen behaviors, e.g., whether the system involves oxygen release and surface reactions, whether the O and TM redox are mised, etc..

For a demonstration of the direct probe of TM-3*d* states through mRIXS, Fig. 1c-e shows the contrast between the Mn-*L* mRIXS-iPFY analysis and conventional XAS. The principle of iPFY has been explained before,[41] and demonstrations of Mn-*L* iPFY analysis of mRIXS were also realized in our high-effeciency mRIXS system.[30] In brief, with the excitation energy range covering Mn-*L* edges, e.g., 638-646 eV for Mn-*L₃* edge, the incident X-rays also excite O-*K* features at emission energy range of 495-510 eV (Fig. 1d). The intensity of such a non-resonant O-$K$ emission is affected by the absorption coefficient of the incident photons at the Mn-$L$ energy range, leading to (inversed) dips corresponding to the intrinsic Mn-$L$ absorption peaks.[30] Therefore, through the high energy resolution of the mRIXS results, a PFY$_{O-K}$ can be obtained by integrating the mRIXS intensity within the O-$K$ emission range (495-510 eV, Fig. 1d), and its inverse value iPFY = 1/PFY$_{O-K}$ provides a bulk-sensitive (photon-in-photon-out) probe of the intrinsic Mn-L absorption coefficient without any spectral distortion. The Mn-$L$ iPFY spectrum in Fig. 1e (red) displays an exact Mn$^{4+}$ spectrum in the charged electrodes,[48] contrasting the spectra from other conventional XAS channels, including TEY, TFY and PFY. Such Mn-$L$ data in the bulk-probe (probe depth of about 200 nm in Mn-$L$ edges) fluorescence mode naturally enables the quantitative fittings of the bulk TM oxidation states, which have been established for various TM-$L$ edges.[39] Additionally, although mRIXS-iPFY still contains surface signals, when compared with the surface sensitive (10 nm probe depth) TEY spectra, such a non-distorted detection with >100 nm probe depth provides the direct contrasts between the bulk and surface chemistry of the electrodes. Below, all Mn-$L$ raw data were collected through full range mRIXS, but for clarify, we focus on the final iPFY results, and their comparison with the TEY surface signals.

**Materials and electrochemical properties**

The Na$_{2/3}$Mg$_{1/3}$Mn$_{2/3}$O$_2$ material is synthesized by poly (vinyl pyrrolidone) (PVP) - assisted gel combustion method.[49,50] This stoichiometry is adjusted so that the nominal valence of Mn in the material is pure +4, i.e., Mn redox is intentionally suppressed during the initial charge. Structural characterizations through XRD and SEM are presented in Supplementary Fig. S1 and S2. The XRD (Fig. S1) shows the same pattern as Na$_{2/3}$[Mg$_{0.28}$Mn$_{0.72}$]O$_2$.[43] All diffraction peaks can be indexed as a hexagonal structure with a $P6_3/mmc$ space group, i.e., P2-type. The peak at 20.6° is due to the formation of $\sqrt{3}$a × $\sqrt{3}$a-type in-plane superlattice between Mg$^{2+}$ and Mn$^{4+}$.[51] Fig. S2 shows the SEM image of the well-crystallized morphology of Na$_{2/3}$Mg$_{1/3}$Mn$_{2/3}$O$_2$ as well as calculated particle size distribution. The particle size is between 0.5 μm and 3 μm. The particles are smooth-faced without secondary-particle structure. Such morphology is generally desirable for improving the electrochemical performance.

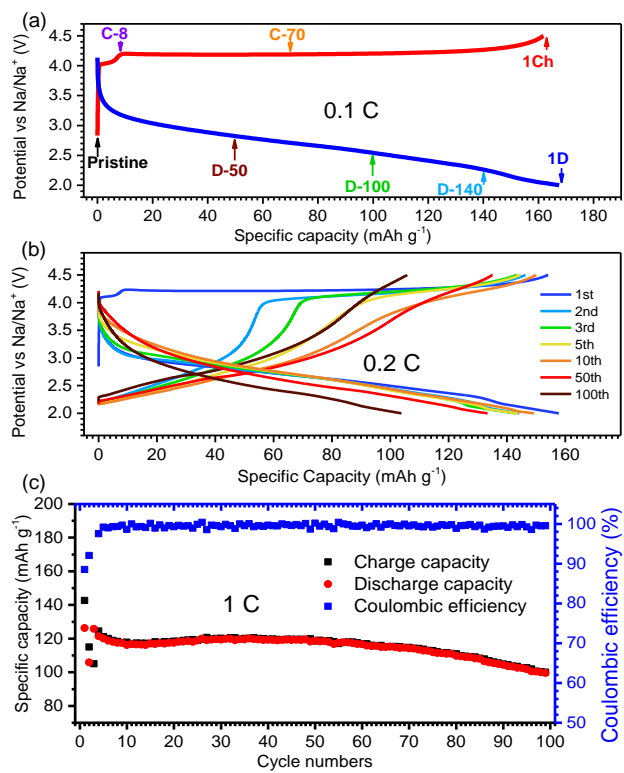

**Fig. 2 | Electrochemical properties of Na$_{2/3}$Mg$_{1/3}$Mn$_{2/3}$O$_2$.** (a) The initial charge and discharge profiles of the Na$_{2/3}$Mg$_{1/3}$Mn$_{2/3}$O$_2$ at 0.1 C. The points that tested by XAS and RIXS are marked with arrows. (b) The electrochemical profiles of different cycles at 0.2 C. (c) Capacity retention and Coulomb efficiency with 1 C cycling rate. See also Figure S3-S7.

The electrochemical performance of Na$_{2/3}$Mg$_{1/3}$Mn$_{2/3}$O$_2$ electrodes is shown in Fig. 2. A short plateau of only 8 mAh g$^{-1}$ capacity appears at 4.0 V before the potential reaches a very long 4.2 V (OCV is ~4.1 V, Fig. S3b) plateau, corresponding to a strong peak at 4.3 V in the cyclic voltammetry curves shown in Fig. S3a. The overall initial charge capacity reaches 162 mAh g$^{-1}$, which is dominated by the 4.2 V plateau. Note again that this improved initial charge capacity is practically important for NIBs. This also means the Na$^+$ can be extracted near 0.56 mol and the formula of charged electrode is Na$_{0.11}$Mg$_{1/3}$Mn$_{2/3}$O$_2$. Although this work does not focus on the material itself, we note this is the highest initial charge capacity value reported for air-stable P2-type sodium transition metal oxides. The good rate capability is shown in Fig. S4. Compared with the initial study of Na$_{2/3}$[Mg$_{0.28}$Mn$_{0.72}$]O$_2$ electrode with 150 mAh g$^{-1}$ charge capacity,[43] the capacity before the 4.2 V charge plateau decreases due to the further suppression of the Mn redox in Na$_{2/3}$Mg$_{1/3}$Mn$_{2/3}$O$_2$. Note that the suppression of the cation TM redox leads to a counterintuitive increase of the initial charge capacity through the long 4.2 V charge plateau. This indirectly indicates the highly active oxygen in the Na$_{2/3}$Mg$_{1/3}$Mn$_{2/3}$O$_2$ electrode during electrochemical operations, providing an optimized candidate for oxygen redox studies.

Fig. 2b and Fig. S5 compare the charge and discharge profiles at 1st, 2nd, 3rd, 5th, 10th, 50th, and 100th cycles. Cycling profiles with other potential range were measured and plotted in Fig. S6. The cycling profile clearly shows that the capacity of the 4.2 V plateau drops continuously with electrochemical cycles. Fig. 2c shows the charge and discharge capacity retention and the coulombic efficiency at 1 C. The capacity retention is 80% after 100 cycles. The average coulombic efficiency during cycling is 99.3% excluding some abnormal point. The long-term cycling performance for 2.0-4.5 V and 1.5-4.5 V at 1 C rate and 2.0-4.5 V at 0.2 C are shown in Fig. S7. The high-voltage plateau is often used as an indicator of oxygen redox. If such assumption were valid, the cycling profile implies that oxygen redox is only partially reversible for about 10 cycles before it disappears in this system. It is therefore critical to experimentally clarify the activity of oxygen redox beyond 10 cycles.

**Mn states in bulk and on surface upon electrochemical cycling**

We first clarify the Mn redox activities of $Na_{2/3}Mg_{1/3}Mn_{2/3}O_2$ electrodes in the bulk and on the surface upon electrochemical cycling. As explained above, Mn-3$d$ valence states in the bulk are measured through the Mn-$L$ mRIXS-iPFY spectra. Surface signals are collected through conventional XAS-TEY[38]. Because both data sets represents intrinsic absorption coefficient without lineshape distortion, they could be reliably quantified based on a simple linear combination of the distinct lineshape of $Mn^{2+/3+/4+}$ references,[48,52] which has been successfully demonstrated for obtaining quantitative contrast of the bulk and surface Mn states.[39,53]

Mn-$L_3$ iPFY and TEY spectra are over-plotted for direct comparison of the bulk and surface Mn states. Spectra with different potentials during the initial cycle are plotted in Fig. 3, and data with extended cycles are displayed in Fig. 4. For the important bulk signal associated with the lattice oxygen redox discussed later, quantitative fittings of iPFY are plotted directly in the figures (dotted line) to show the satisfactory simulation. The full energy range TEY spectra of more samples and their quantitative fitting plots are shown in supplementary Fig. S8 and Fig. S9. The conventional TFY spectra are also plotted in Fig S10 and Fig. S11 for reference. Fittings of both iPFY and TEY results highly agree with experimental data, providing a reliable determination of the Mn oxidation states in bulk and on surface. The contrast between the iPFY and TEY spectra clearly shows the emergence of a surface inactive layer of $Mn^{2+}$ upon cycling, indicated by the 640 eV peak only found in TEY data. Such a contrast, as well as the overall bulk redox evolution upon cycling, is visualized in the fitting results plotted in Fig. 3b and Fig. 4b. The values of Mn valence percentages by fitting iPFY (bulk) and TEY (surface) spectra are shown in table 1.

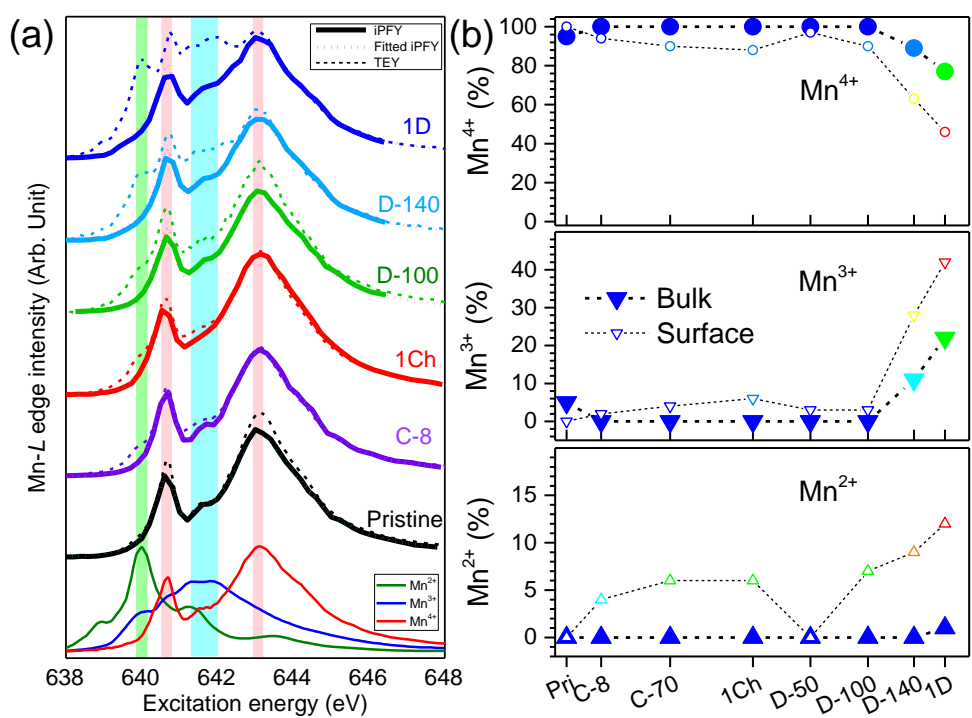

**Fig. 3 | Mn-$L_3$ iPFY and TEY spectra and the quantified bulk and surface Mn valence states** of $Na_{2/3}Mg_{1/3}Mn_{2/3}O_2$ electrodes at representative SOC levels during the initial cycle indicated in Fig. 2a. (a) The Mn-$L_3$ edge iPFY (solid), fitted iPFY (dotted) and TEY (dashed) spectra. References of MnO ($Mn^{2+}$), $Mn_2O_3$ ($Mn^{3+}$) and $Li_2MnO_3$ ($Mn^{4+}$) are plotted at the bottom for direct comparison. (b) The fitting results of the Mn valence distributions in bulk (solid points) and on surface (hollow points) at different electrochemical stages. Error bars for fitted bulk Mn states are smaller than the symbols and are therefore not shown. See also Figure S8 and S10.

For the initial charge and discharge cycle (Fig. 3), a trace amount of about 5% $Mn^{3+}$ is found in the pristine material, which is fully oxidize to $Mn^{4+}$ during the very short 8 mAh $g^{-1}$ charging process (C-8), corresponding to the small charge capacity at 4.0 V in Fig. 2a. The $Mn^{4+}$ state then maintains unchanged until the potential drops below 2.5 V during discharge. It is clear that the adjusted stoichiometry of $Na_{2/3}Mg_{1/3}Mn_{2/3}O_2$ indeed suppresses the Mn redox during the initial charge process, and Mn in bulk remains at $Mn^{4+}$ throughout the large-capacity 4.2 V charge plateau. However, $Mn^{3+}$ starts to develop at low potentials during discharge (D100-D140). The D-140 and fully discharged (1D) electrode have about 11% and 23% $Mn^{3+}$ with a trace contribution from $Mn^{2+}$, corresponding to 0.07 and 0.16 mol electron transfer per 1 mol $Na_{2/3}Mg_{1/3}Mn_{2/3}O_2$ due to Mn reduction, respectively.

In contrast, the Mn states on the surface show a very different trend. The surface Mn of the pristine electrode is fully $Mn^{4+}$, suggesting that $Mn^{3+}$ does not appear on the surface of $Na_{2/3}Mg_{1/3}Mn_{2/3}O_2$ like the spinel $LiMn_2O_4$. However, during charging, the surface $Mn^{2+}$ and $Mn^{3+}$ are increasing, which is different from that of $Na_{0.44}MnO_2$ in which no oxygen redox is involved.[53] At the beginning stage of discharge, the surface Mn is oxidized slightly to +4 valence (D-50) and then gradually reduced during the discharge progress, resulting in a rapid increase of surface $Mn^{2+}$. Therefore, the surface activities in such an oxygen redox system are obviously affected by two different factors: some surface reactions that are likely associated with the oxidized oxygen during charge, and the surface reduction effect by the electrochemical discharging itself.

For the electrodes with extended cycles (Fig. 4), the $Mn^{2+}$ on the surface continuously increases with the cycle. The surface $Mn^{2+}$ content reaches 35% after 100 cycles, which could be seen directly through the TEY spectra of both charged and discharged electrodes. This indicate that only part of the surface $Mn^{2+}$ is electrochemically active. On the other hand, the iPFY analysis of the Mn states in the bulk shows that $Mn^{3+/4+}$ redox remains stable over a hundred cycles, despite of the obvious change of the electrochemical profile (Fig. 2b). Although the $Mn^{3+/4+}$ redox is stable upon cycling, the amount of $Mn^{3+}$ displays very slow increment upon cycling. We note that the TM valence drop in the bulk has recently been reported in Li-rich materials to explain the voltage fade there.[44] Mn-*L* edge provides a more direct quantification of the *3d* valence states in the bulk and on the surface, and the results here indicate voltage fade in $Na_{2/3}Mg_{1/3}Mn_{2/3}O_2$ could also be from the decreasing Mn valence upon cycling, especially on the electrode surface. Of course, the voltage fade here is smaller than that in typical Li-rich compounds.[44] Future detailed and comparative studies on voltage fade based on the same type of materials are critical to further clarify the origin of voltage decay and its association with redox activities.

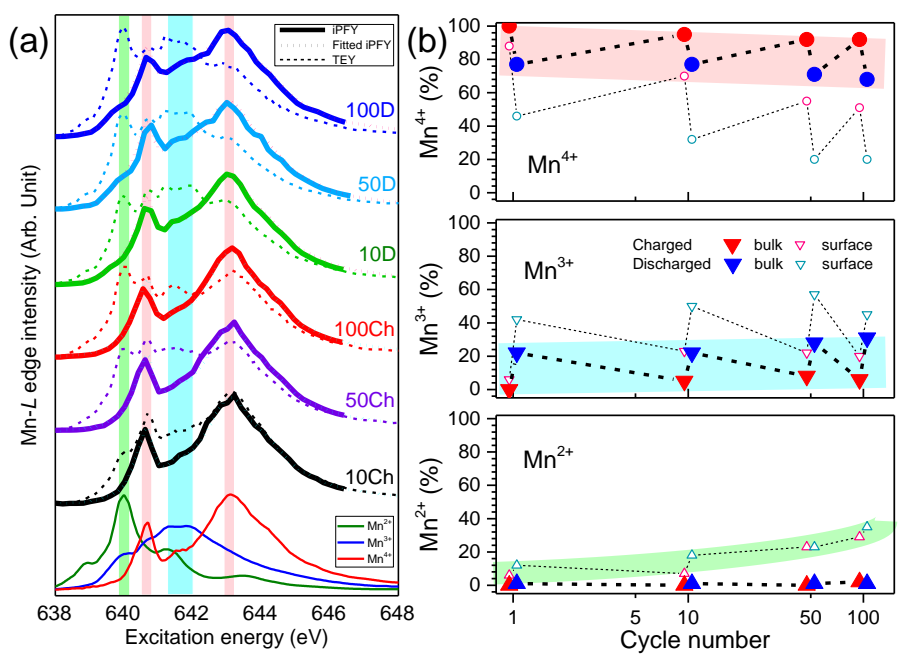

**Fig. 4 | Mn-*L₃* edge spectra and the calculated bulk and surface Mn valence**

**states** of charged and discharged $Na_{2/3}Mg_{1/3}Mn_{2/3}O_2$ electrodes after different cycles. (a) The Mn-$L_3$ edge iPFY (solid), fitted iPFY (dotted) and TEY (dashed) spectra of $Na_{2/3}Mg_{1/3}Mn_{2/3}O_2$ electrodes. References of MnO ($Mn^{2+}$), $Mn_2O_3$ ($Mn^{3+}$) and $Li_2MnO_3$ ($Mn^{4+}$) are plotted at the bottom for direct comparison. (b) The fitting results of the Mn valence distributions in bulk (big solid points) and on surface (small hollow points) at different cycles. Error bars for fitted bulk Mn states are smaller than the symbols and are therefore not shown. See also Figure S9 and S11.

**Table 1.** Mn valence percentage by fitting from iPFY (bulk) and TEY (surface) spactra. The error is below ± 3 percentage estimated from the uncertainty in the fitting procedure.

| Sample | Bulk | | | Surface | | |
|---|---|---|---|---|---|---|
| | $Mn^{2+}$ | $Mn^{3+}$ | $Mn^{4+}$ | $Mn^{2+}$ | $Mn^{3+}$ | $Mn^{4+}$ |
| Pri | 0 | 5 | 95 | 0 | 0 | 100 |
| C-8 | 0 | 0 | 100 | 4 | 2 | 94 |
| C-70 | - | - | - | 6 | 4 | 90 |
| 1Ch | 0 | 0 | 100 | 6 | 6 | 88 |
| D-50 | - | - | - | 0 | 3 | 97 |
| D-100 | 0 | 0 | 100 | 7 | 3 | 90 |
| D-140 | 0 | 11 | 89 | 9 | 28 | 63 |
| 1D | 1 | 22 | 77 | 12 | 42 | 46 |
| 10Ch | 0 | 5 | 95 | 7 | 23 | 70 |
| 50Ch | 0 | 8 | 92 | 23 | 22 | 55 |
| 100Ch | 2 | 6 | 92 | 29 | 20 | 51 |
| 10D | 1 | 22 | 77 | 18 | 50 | 32 |
| 50D | 1 | 28 | 71 | 23 | 57 | 20 |
| 100D | 1 | 31 | 68 | 35 | 45 | 20 |

**Quantify the Oxygen redox reversibility during initial cycles**

While the Mn-$L$ iPFY suggests that majority of the capacity of $Na_{2/3}Mg_{1/3}Mn_{2/3}O_2$ electrodes is not from Mn redox, it only provides an indirect evidence on oxygen redox, especially with the clear evidence on strong surface reactions discussed above. Direct experimental evidences and quantifications of the oxygen state evolution upon extended cycles remain critical and are the key contributions of this work. The conventional O-$K$ XAS (TEY and TFY) results of representative samples are shown in supplementary Fig. S12. As discussed in the introduction session, XAS pre-edge lineshape changes do not represent the oxygen redox in battery electrodes, instead, it represents the change on TM-O hybridization strength, and the changes upon electrochemical cycling could thus be observed in almost all battery electrodes, including spinel and olivine materials without oxygen redox.[20,42,54] Therefore, features in the 528-534 eV pre-edge range needs to be further resolved to extract the reliable signals of oxygen redox states for quantifications.

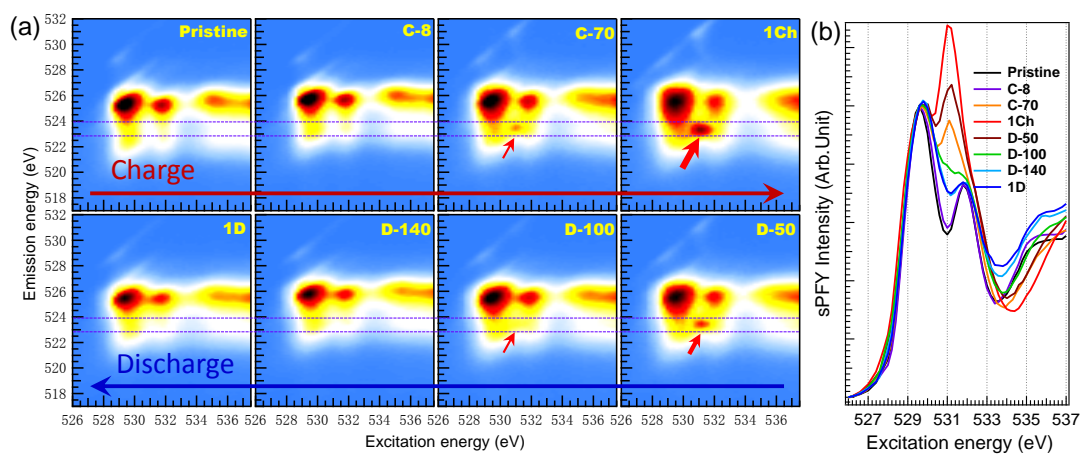

**Fig. 5 | O-*K* mRIXS and sPFY during initial cycle.** (a) The mRIXS images of Na$_{2/3}$Mg$_{1/3}$Mn$_{2/3}$O$_2$ electrodes at different SOC during initial cycle. Samples are marked in Fig. 2(a). The key oxygen redox feature that emerges at high potentials is indicated by the red arrows. (b) sPFY spectra extracted from mRIXS by integrating the characteristic 523.7 eV emission energy range, as indicated by the two horizontal purple dashed lines (see technical demonstrations of sPFY in Fig. 1). See also Figure S12.

Further resolving the XAS features in the 528-534 eV range is exactly what mRIXS measures.[20] Fig. 5a show the O-*K* mRIXS of a series of electrodes, again denoted in Fig. 2a, at different SOC during the initial charge and discharge. As specified above, the broad features around 525 eV emission energy (vertical axis) correspond to the O$^{2-}$ states in typical TM oxides,[45] which can be seen in all samples. These features get broadened in lineshape and enhanced in intensity upon charging (oxidation) process, consistent with the many XAS observations but irrelevant to oxygen redox states. In contrast, the feature at 523.7 eV emission energy (red arrows) starts to emerge only when the potential reaches the 4.2 V plateau (Fig. 2a, after C-8), and becomes strong at the charged state. During discharge, this feature gets weakened gradually and disappears when the potential is below 2.5 V (after D-100).

Fig. 5b is the sPFY of the 523.7 eV emission-energy feature extracted from mRIXS images, and the integrated area (the calculation is explained in method part) of the characteristic 531 eV sPFY peak is displayed in Fig. 6 (hollow circles) together with the cycling profile. The 531 eV sPFY peak emerges after the 4.2 V charge plateau starts, reaches the maximum intensity at the fully charged state, and gets weakened until about 2.5 V in the discharge process.

Combining the area of the 531 eV O-*K* mRIXS-sPFY peak with the quantitative results of bulk Mn redox obtained through Mn-*L* mRIXS-iPFY (Fig. 3), we now could explain the total capacity with the independent Mn and O redox quantifications (Fig.

6). Overall, changes on Mn-$L$ (Fig. 3) and O-$K$ (Fig. 5) spectroscopic results take place at different electrochemical stages in Na$_{2/3}$Mg$_{1/3}$Mn$_{2/3}$O$_2$ system. The O-$K$ results show that oxygen oxidation starts at the 4.2 V during initial charge, and stops at 2.5 V during the following discharge, within which, Mn-$L$ displays no change. This clearly defines the crossover points from oxygen to Mn redox reactions at 4.2 V during charge and 2.5 V during discharge. Therefore, the corresponding charge compensation of Mn (blue in Fig. 6a) and O (red in Fig. 6a) redox are 0.03 mol and 0.53 mol during charging, and 0.16 mol and 0.42 mol during discharge from electrochemical experiments (x axis of Fig. 6).

These charge compensation numbers from the electrochemical measurements could now be determined by our Mn-$L$ mRIXS-iPFY and O-$K$ mRIXS-sPFY quantification values. First, for the main Mn redox contribution during discharge, the quantified electron transfer number from Mn redox by calculations based on the Mn valences from mRIXS-iPFY results (Fig. 3 and Table 1) is 0.16 mol at discharged state per 1 mol Na$_{2/3}$Mg$_{1/3}$Mn$_{2/3}$O$_2$ (right-most column bar), matching exactly the electrochemical Na intercalation value during discharge (0.16 mol on bottom x axis). Second, for O redox contributions, the O-$K$ mRIXS-sPFY 531 eV peak intensity (open circles in Fig. 6a) displays a changing ratio of 0.77/0.61 of the charge/discharge process, which matches exactly the electrochemical charge transfer of the oxygen redox regimes (0.53/0.42, bottom x axis of Fig. 6a).

If we assume a linear dependence between the amount of O redox and O-$K$ mRIXS-sPFY 531 eV peak area changes, the value of the spectroscopic peak area changes (left vertical axis of Fig. 6b) are in great agreement with the electrochemical electron transfer numbers (x axis of Fig. 6b). We note that, because the sPFY spectra were normalized to the hybridization feature for comparison purpose (see Method) and the hybridization feature is typically enhanced at charged state, the assumption of the linear dependence above should be modified for an absolute quantification of the amount of O redox at each electrochemical state (underestimated with linear dependence). However, in this work, we focus on the variation of the sPFY peak area to address the reversibility of O redox, which only requires the same hybridization level at the same electrochemical state. The great agreement between sPFY analysis and electrochemistry here suggests this is a reasonable condition at least for Na$_{2/3}$Mg$_{1/3}$Mn$_{2/3}$O$_2$, where hybridization enhancement upon charge is likely minimized through mostly oxidized oxygen, instead of strong TM redox in other systems. Furthermore, our quantification here is based on a self-consistent analysis of both the spectroscopic variation and electrochemical profile. The independent quantification values of Mn and O redox reactions match and decipher the electrochemical capacity. The results from both sPFY peak area changes and electrochemical capacity (with range defined by mRIXS as explained above) reveal directly the reversibility rate of the lattice oxygen redox during the initial cycle of $Na_{2/3}Mg_{1/3}Mn_{2/3}O_2$, i.e, 0.61/0.77 = 79%.

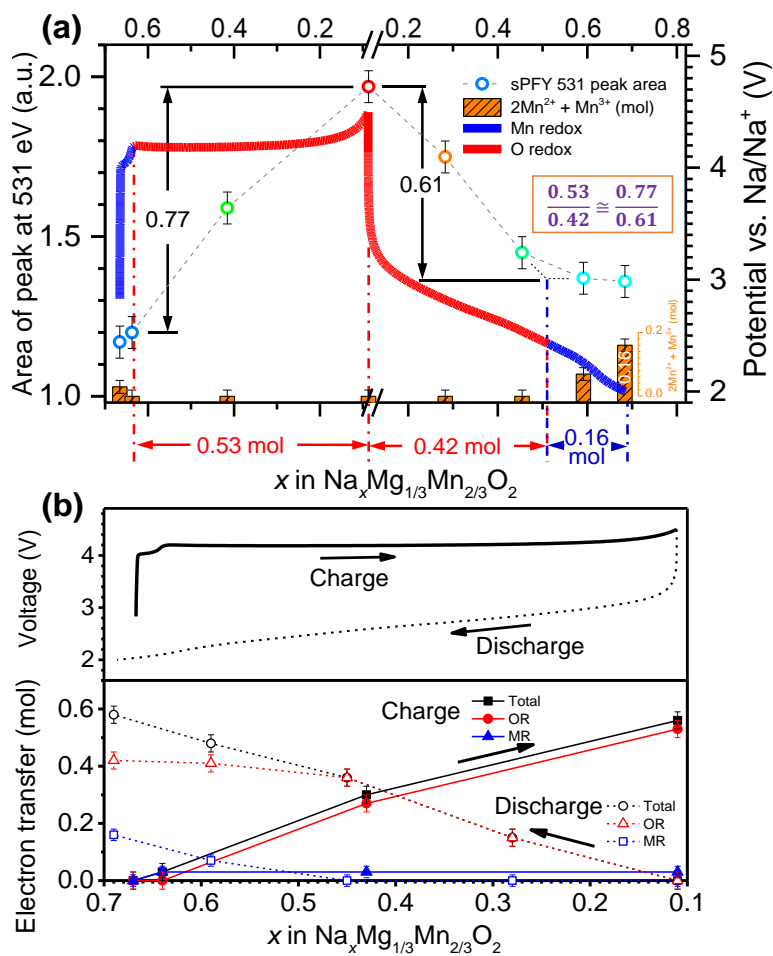

**Fig. 6 | Decipher the total electrochemical capacity by independent quantifications of Mn redox and changing O redox states.** (a) The blue and red color of the electrochemical profile represent Mn redox and O redox regimes. The crossover points of 4.2 V during charge and 2.5V during discharge are determined experimentally by whether the Mn and O spectroscopic signals change. In particular, the O redox contributions are 0.53 mol and 0.42 mol during charge and discharge, respectively. Hollow circles are areas of O-*K* mRIXS-sPFY 531 eV peaks. The change of the peak area is 0.77 during charge and 0.61 during discharge. The ratio of the oxygen redox contributions during charge/discharge are consistently determined by either the electron transfer numbers (0.42/0.53) or mRIXS-sPFY peak area (0.61/0.77) as 79%. The Mn redox contributions are quantified by the Mn valences from Mn-*L* mRIXS-iPFY results (columns bars on the bottom). For example, the fully discharged state includes 0.16 mol electron transfer from Mn redox, which matches exactly the regime of Mn redox in electrochemical profile (0.16 mol). (b) Accumulated electron transfer of Mn redox and O redox during charging and discharging. Again, Mn redox contributions are calculated based on Mn valences from mRIXS-iPFY. O redox is calculated based on O-*K* mRIXS-

sPFY peak area changes, assuming a linear dependence between the amount of O redox and the sPFY 531 eV peak area (see text on assumptions). Error bars are over-estimated here considering all possible contributions from uncertainty in the fitting procedure, the resolving power of RIXS, and the upper and lower limits of scaling.

### Cyclability of oxygen redox upon extended cycles

The characteristic oxygen-redox feature in mRIXS and the quantitative peak area analysis through sPFY empower us to quantify the oxygen redox cyclability upon electrochemical cycling. This approach is employed to study electrodes with extended cycles, as shown in Fig. 7. In general, the oxygen-redox feature in mRIXS images displays amazing cyclability over a hundred cycles of $Na_{2/3}Mg_{1/3}Mn_{2/3}O_2$ (Fig. 7a), with a strong intensity at charged state (red arrows in top row) and disappearance at the discharged state (bottom row). Fig. 7b are again the mRIXS-sPFY spectra around 523.7 eV emission energy. The amount of O redox at different cycles are quantified by the sPFY 531 eV peak intensity differences between the charged and discharged states at the 1st, 10th, 50th, and 100th cycles. The summary of the amount of Mn redox, O redox, and the O-$K$ mRIXS-sPFY 531 eV peak area change up to 100 cycles are displayed in Fig. 7c and Table 2.

**Table 2. Charge transfer numbers from $Na^+$, Mn redox (MR) and O redox (OR) upon electrochemical cycling.** The electron transfer number of MR, $n_{MR}$, is calculated by the Mn valence change from charged to discharged state (Fig. 4 and Table 1) multiplied by 2/3 mol (2/3 mol Mn per 1 mol $Na_{2/3}Mg_{1/3}Mn_{2/3}O_2$). The electron transfer number of OR, $n_{OR}$, is calculated by the mRIXS-sPFY 531 eV peak area difference between charged and discharged state (Fig. 7b, c), multiplied by the factor of (0.42/0.61) mol from the first discharge, as discussed in detail before (Fig. 6a).

| Cycle number | $n_{Na}$ in discharge (mol) | Mn average valence | | $n_{MR}$ (mol) | $n_{OR}$ (mol) |
| --- | --- | --- | --- | --- | --- |
| | | Charged | Discharged | | |
| 1 | 0.58 | 4 | 3.76 | 0.16 | 0.42 |
| 10 | 0.56 | 3.95 | 3.76 | 0.13 | 0.44 |
| 50 | 0.53 | 3.92 | 3.70 | 0.15 | 0.45 |
| 100 | 0.47 | 3.90 | 3.67 | 0.15 | 0.39 |

The quantitative analysis here shows that the oxygen redox activities are very stable up to 50 cycles, then start to decay with about 87% of the oxygen redox sustained in the 100th cycle (Fig. 7c and Table 2). We note that such a quantification with extended cycles is impossible through the conventional XAS even technically, as shown in Fig. S12, because the XAS spectra of charged and discharged electrodes become almost identical after only tens of cycles. An interesting phenomenon is that Fig. 7b shows clearly that the 531 eV sPFY peak area at charged state does not decrease with cycling, however, the overall intensity of the fully discharged states keeps increasing upon cycling. While such an increase is related to the overall peak broadening from structural amorphization upon cycling, i.e., lifting the dip between two peaks, the very systematic change at discharged states also implies that the decreasing of oxygen redox contents beyond 50 cycles may be associated with the deep discharging.

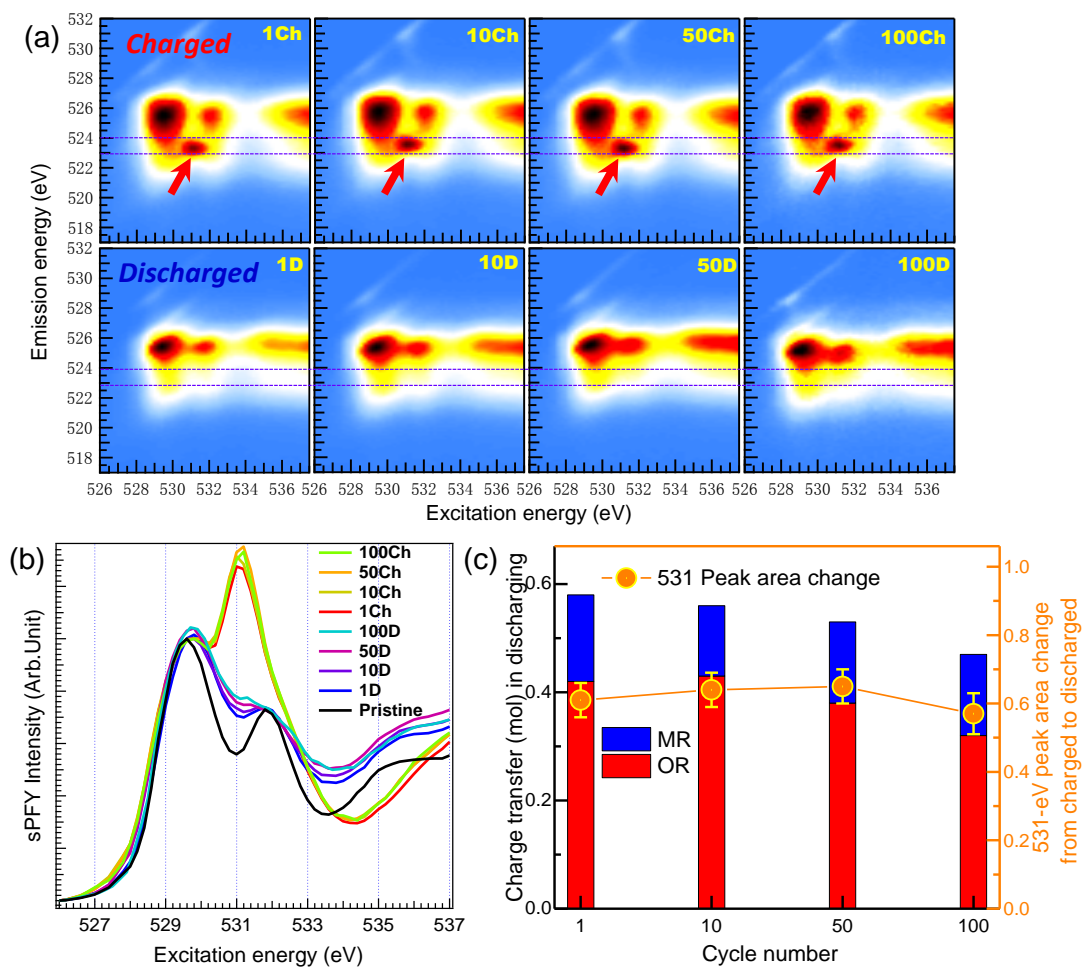

**Fig. 7 | Cyclability of oxygen and Mn redox upon cycling.** (a) O-*K* mRIXS images of a series of charged (top) and discharged (bottom) electrode after the 1[st], 10[th], 50[th] and 100[th] cycles. Red arrows indicate the oxidized oxygen features in charged states. (b) sPFY spectra extracted from mRIXS by integrating the characteristic 523.7 eV emission energy range (purple dashed lines). The oxidized oxygen peak remains strong and reversible over a hundred cycles. (c) The quantified electron transfer of Mn redox (blue bar), the expected value of O redox contribution (red bar) calculated by subtracting Mn redox contribution from total capacity, and the 531 O-*K* sPFY peak area changes between charged and discharged states at extended cycles. Error bars are over-estimated here considering all possible contributions from uncertainty in the fitting procedure, the resolving power of RIXS, and the upper and lower limits of scaling. See also Figure S12.

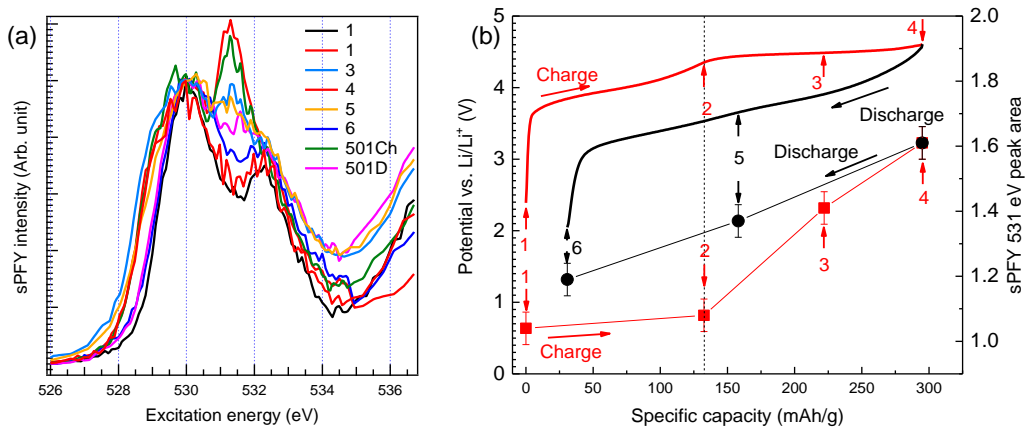

**Fig. 8 | O-*K* mRIXS-sPFY quantifications of lattice oxygen redox in Li$_{1.17}$Ni$_{0.21}$Co$_{0.08}$Mn$_{0.54}$O$_2$.** (a) The mRIXS-sPFY at 523.7 eV emission energy range (see Fig. 1a) of electrodes at different SOC during initial and extended cycle based on mRIXS results reported previously.[31]. Samples during initial cycle are marked on the electrochemical profile in (b). Red squares in (b) are integrated area (right vertical axis) of the characteristic mRIXS-sPFY peak at 531 eV during charging (red arrows), which increases during the charge plateau but not below. Black dots represent the integrated oxygen-redox peak area upon discharge (black arrows). Integrated peak area values after extended cycles are also analyzed and discussed in text, but not shown in (b) for display clarity. Error bars are over-estimated here considering all possible contributions from uncertainty in the fitting procedure, the resolving power of RIXS, and the upper and lower limits of scaling.

## mRIXS Quantifications of the lattice oxygen redox in Li$_{1.17}$Ni$_{0.21}$Co$_{0.08}$Mn$_{0.54}$O$_2$

In this session, we demonstrate that the mRIXS-sPFY analysis here could be employed for quantitative studies of lattice oxygen redox in various battery electrodes, including those with oxygen gas evolutions. It is important to emphasize that what mRIXS detects is the specific lattice oxygen redox within the bulk electrodes, i.e., not the general/total oxygen redox that includes other irreversible oxygen involvements. We also note that the lattice oxygen redox is the critical part of the reactions that could be potentially useful for battery devices. mRIXS technique, as a photon-in-photon-out bulk probe under vacuum environment, provide unique opportunities to focus on this practically meaningful lattice oxygen redox reactions, and to quantify its reversibility and cyclability.

For a comparative study, we employed the same O-*K* mRIXS-sPFY quantification to a more complex Li-rich compound, Li$_{1.17}$Ni$_{0.21}$Co$_{0.08}$Mn$_{0.54}$O$_2$. Oxygen redox in Li-rich electrode system has been extensively studied recently.[5,12,13,28,33,44] We note that STXM results through non-distorted transmission mode could quantify the oxygen redox during the initial cycle,[31] but samples for STXM need to be processed into individual nano-particles due to the limited penetration depth of soft X-rays. In comparison, mRIXS could be directly employed on any form of electrodes with any size of particles and at any electrochemical states after extended cycles.

Fig. 8a shows the mRIXS-sPFY results of a series of $Li_{1.17}Ni_{0.21}Co_{0.08}Mn_{0.54}O_2$ electrodes indicated on the electrochemical profile in Fig. 8b, by integrating the same energy window as for our $Na_{2/3}Mg_{1/3}Mn_{2/3}O_2$ analysis (Fig. 1a) of the mRIXS results reported previously.[31] Although a bit noisier than that of $Na_{2/3}Mg_{1/3}Mn_{2/3}O_2$, the 531 eV peak intensity displays clear variation at different electrochemical states. The peak area is plotted together with the electrochemical profile in Fig. 8b (offset for display clarity). It is clear that the peak area remains unchanged before the 4.5 V charge plateau, but starts to increase during the charge plateau with roughly a linear dependence on charge capacity. In the following discharge, the area value decreases through most of the discharge process but did not recover to the pristine state. The change of the 531 eV peak area in the initial charging and discharging is 0.55 and 0.42, respectively; therefore, the reversibility of the oxygen redox in the initial cycle is about 76%. After 500 cycles, the 531 eV peak area change from charge to discharged states is 0.24, suggesting a cyclability of 44%.

There are several important contrasts between the $Li_{1.17}Ni_{0.21}Co_{0.08}Mn_{0.54}O_2$ and $Na_{2/3}Mg_{1/3}Mn_{2/3}O_2$ systems indicated by our quantification results. First, there is a clear crossover point between the O and Mn redox reactions in $Na_{2/3}Mg_{1/3}Mn_{2/3}O_2$ (Fig. 6a and discussions above), therefore, capacity contributions from oxygen redox could be well defined during both charge and discharged process that match the spectroscopic quantifications. However, for $Li_{1.17}Ni_{0.21}Co_{0.08}Mn_{0.54}O_2$, oxygen redox reactions take place throughout the discharge process mixed together with TM redox reactions.[31,44] Therefore, because of the contributions from TM redox, the electrochemical discharge capacity is much larger than that of the oxygen-redox region during charge, i.e., the charge plateau. Nonetheless, mRIXS-sPFY is able to distinguish the lattice oxygen redox contributions from such a mixed redox process and reveals the 76% initial cycle reversibility as elaborated above.

Second, another important contrast emerges if we further examine the quantification values of the oxygen redox contributions during the initial charge process. The electron transfer in the 4.5 V charging plateau for $Li_{1.17}Ni_{0.21}Co_{0.08}Mn_{0.54}O_2$ is 0.52 mol, with the change of mRIXS-sPFY peak area of 0.55 (Fig. 8b). $Na_{2/3}Mg_{1/3}Mn_{2/3}O_2$ displays 0.77 peak area change with 0.53 mol electron transfer (Fig. 6a). Although the absolute value of the conversion ratio between the variation of mRIXS-sPFY peak area and electron transfer number could be highly material dependent due to the different hybridization background, the $Li_{1.17}Ni_{0.21}Co_{0.08}Mn_{0.54}O_2$ obviously displays much lower mRIXS variation than that in $Na_{2/3}Mg_{1/3}Mn_{2/3}O_2$ upon the same amount of electrochemical charge transfer. This is again expected because mRIXS detects only the lattice oxygen redox, without contributions from the oxygen release and surface reactions that has been reported in many previous works for Li-rich electrodes.[28,35,55-58] As a very rough estimation, if we ignore the different TM-O hybridization background in the two systems that affect the absolute sPFY intensity of the 529.7 eV peak after scaling (see Methods), lattice oxygen redox contribution could be estimated to be about 70% (0.55/0.77) of the total capacity during the initial charge plateau with about 0.38 mol electron transfer. Other capacity contributions mainly including oxygen release thus lead to 0.52-0.38=0.14 mol electron transfer, corresponding to about 40 mAh g$^{-1}$. This is surprisingly consistent with the measured capacity contribution from oxygen release in $Li_{1.2}Ni_{0.2}Mn_{0.6}O_2$ (35 mAh g$^{-1}$) in our previous work,[33] considering other surface reactions also contribute to the charge capacity in Li-rich systems.[28,59-61] For the same reason that mRIXS detects only the lattice oxygen redox, although the 76% reversibility seems surprisingly high for a Li-rich system, it is reasonable as this is from distinguished lattice oxygen redox only. The capability of mRIXS technique to distinguish lattice oxygen redox from other irreversible oxygen contributions delivers an important information that lattice oxygen redox itself is not as irreversible as we commonly believe, i.e., the irreversibility of the general oxygen redox discussion is largely from non-lattice oxygen redox reactions, not lattice oxygen redox reactions.

Finally, the good match between spectroscopic quantifications and electrochemical capacity for $Na_{2/3}Mg_{1/3}Mn_{2/3}O_2$ indicates this system is dominated by stable lattice oxygen redox with limited non-lattice oxygen contributions. This allows reasonable quantifications of the absolute charge transfer numbers based on the mRIXS-sPFY peak area changes (Fig. 7c). It is important to note, however, such absolute quantification could/should only be performed for systems with negligible non-lattice oxygen redox reactions, and the conversion ratio between peak area and electron transfer could be material dependent. In a complex system with oxygen release and other oxygen contributions, the peak area change reflects only the lattice oxygen redox contribution part. Therefore quantitative comparison with electrochemical capacity is relatively more reliable during discharging because most non-lattice oxygen reactions, such as oxygen release, take place during high-voltage charging. Nonetheless, mRIXS-sPFY quantification successfully distinguishes the contributions from lattice oxygen redox reactions to the electrochemical capacity, and reveals the intrinsic behavior of lattice oxygen redox in the Li-rich systems. Further works remain necessary to uncover the different redox behaviors in many other electrode systems, but the quantifications of the two systems here demonstrate the effectiveness of the mRIXS-sPFY to selectively probe the critical lattice oxygen redox reactions.

**Summary and Conclusions**

This work firstly introduces two analysis methods to quantify the bulk anionic and cationic redox through the recently developed mRIXS technique. Through the sPFY quantification of the characteristic oxygen redox feature in O-$K$ mRIXS, we demonstrate a reliable quantification of the reversibility of lattice oxygen redox. Through the iPFY extracted from Mn-$L$ mRIXS, bulk signals with non-distorted lineshape could be directly fitted for quantitative values of Mn oxidation state concentrations. The O and Mn redox are thus identified independently and consistently interpret the total electrochemical capacity. More importantly, the reversibility of lattice oxygen redox during the initial cycle, and the cyclability over extended cycles, could be quantified through such photon-in-photon-out bulk probes.

We synthesize a P2-Type $Na_{2/3}Mg_{1/3}Mn_{2/3}O_2$ NIB material with nominally only $Mn^{4+}$. Counterintuitively, such an adjustment improves the initial charge capacity to 162 mAh g$^{-1}$, higher than the capacity with other stoichiometry in this family. Quantitative analysis of Mn-$L$ mRIXS-iPFY shows that only 8 mAh g$^{-1}$ capacity is contributed by the trace amount of $Mn^{3+}/Mn^{4+}$ redox during the initial charge. Almost all the improved charge capacity is from a 4.2 V high-potential plateau with $Mn^{4+}$ throughout the process. In the meantime, the O-$K$ mRIXS-sPFY quantification reveals the contrast of the amount of oxygen redox during the initial charge and discharge process, which matches the amount of emerging Mn redox and deciphers the Mn and O redox contributions to the total capacity: the improved charge capacity through the 4.2 V plateau is purely from oxygen redox reactions. During discharge, the start 0.42 mol Na intercalation is compensated by oxygen redox reaction and then 0.16 mol Na intercalation is accompanied by Mn redox, in which $Mn^{4+}$ is reduced to $Mn^{3+}$. The Mn and O redox are well separated in different potential regions in $Na_{2/3}Mg_{1/3}Mn_{2/3}O_2$, as clearly defined by mRIXS data, and the oxygen redox in the initial cycle is 79% reversible.

Over extended cycles, the oxygen redox in $Na_{2/3}Mg_{1/3}Mn_{2/3}O_2$ is highly cyclable until the 50[th] cycle, and 87% of the oxygen redox sustains at the 100[th] cycles, regardless of the disappearance of the charge plateau after only 10 cycles. Such a dissociation between the high voltage plateau and oxygen redox was also observed in Li-rich compounds after initial cycle,[31] but now clearly revealed in NIB systems too.[20]

The mRIXS-sPFY quantification established in this work addresses a critical parameter of lattice oxygen redox: the reversibility. The quantification value of $Li_{1.17}Ni_{0.21}Co_{0.08}Mn_{0.54}O_2$ system reveals a decent initial-cycle reversibility of 76% and 44% retention after 500 cycles of the lattice oxygen redox. However, compared with $Na_{2/3}Mg_{1/3}Mn_{2/3}O_2$, the spectroscopic variation from lattice oxygen redox in $Li_{1.17}Ni_{0.21}Co_{0.08}Mn_{0.54}O_2$ is much lower with the same electrochemical charge compensation, due to other oxygen contributions, such as oxygen release and surface reactions. Therefore, mRIXS-sPFY analysis successfully distinguishes the lattice oxygen redox from other capacity contributions in complex systems with various redox-active oxygen activities.

This work establishes a reliable benchmark and methodology for quantifying the reversibility of lattice oxygen redox in bulk electrodes. Being able to distinguish and quantify the lattice oxygen redox from other oxygen contributions through mRIXS enables analysis of the intrinsic lattice oxygen redox behaviors. While we focus on quantifying only the reversibility of lattice oxygen redox in this work, more sophisticated mRIXS analysis could potentially reach an absolute amount of oxygen redox contributions based on reliable references and fundamental understandings of mRIXS-sPFY lineshape, an important technical topic to pursue in the future. Scientifically, direct experimental evidences of the highly reversible and cyclable lattice oxygen redox provides the optimism on improving $3d$ TM based electrode materials for both Li-ion and Na-ion batteries. The combined quantification results of both cationic and anionic redox reactions not only provide direct experimental verifications that electrochemically inactive dopants, e.g., Mg or Li, suppress the TM redox and enables reversible lattice oxygen redox,[62] more importantly, it also shows that mRIXS is powerful to evaluate other and detailed material optimizations towards reversible lattice oxygen redox in future works, e.g., the difference between Mg and Li dopants. Additionally, the similar mRIXS feature in the LIB and NIB electrodes indicates a common oxygen redox mechanism that is yet to be uncovered. The comparison between the two different systems implies the importance of distinguishing lattice oxygen redox from other oxygen activities for revealing the intrinsic properties of the practically meaningful lattice oxygen redox activities.

## EXPERIMENTAL PROCEDURES

**$Na_{2/3}Mg_{1/3}Mn_{2/3}O_2$ Synthesis:** $Na_{2/3}Mg_{1/3}Mn_{2/3}O_2$ was prepared by poly (vinyl pyrrolidone) (PVP) -assisted gel combustion method[49,50]. Stoichiometric $NaOAc \cdot 4H_2O$, $Mg(OAc)_2 \cdot 4H_2O$ and $Mn(OAc)_2 \cdot 4H_2O$, and PVP (the molar ratio of PVP monomer to total metal ions was 2.0) were dissolved in deionized water and pH = 3 was achieved by adding 1:1 $HNO_3$. The mixture was stirred at 120 °C to obtain dried gel. The dried gel was ignited on a hot plate to induce a combustion process which lasted for several minutes. The resulting precursor was preheated at 400 °C for 2 h and then calcined at 900 °C for 6 h with the heating rate of 5 °C $min^{-1}$. After heat treatment, the oven was switched off and the sample was cooled down naturally. The whole process was performed in air. The control of the final stoichiometry is consistent with and directly confirmed by the electrochemical and spectroscopic results throughout this work.

**Morphology and crystal structure characterization:** The morphology was examined using a JEOL 7500F scanning electron microscope (SEM). The analysis of the phase purity and the structural characterization were made by X-ray powder diffraction (XRD) using a Bruker D2 PHASER diffractometer equipped with Cu Kα radiation.

**Electrochemical tests:** The $Na_{2/3}Mg_{1/3}Mn_{2/3}O_2$ cathode was prepared by mixing 80 wt.% active material, 10 wt.% acetylene black (AB) and 10 wt.% polyvinylidene fluoride (PVdF) binder in N-methylpyrrolidone (NMP) to form a slurry. The slurry was doctor-bladed onto aluminum foil, dried at 60 °C, and then punched into electrode discs with a diameter of 12.7 mm. The prepared electrodes were dried at 130 °C for 12 h in a vacuum oven and show typically an active material loading of about 4 mg cm$^{-2}$. The electrochemical cells were fabricated with the $Na_{2/3}Mg_{1/3}Mn_{2/3}O_2$ cathode, sodium foil anode, 1 mol L$^{-1}$ NaClO$_4$ in propylene carbonate (PC) as electrolyte, and double layered glass fiber as separator in an argon-filled glove box. Electrochemical performances were evaluated using CR2325 coin cells. Galvanostatic charge-discharge tests were performed using Maccor 4000.

**$Li_{1.17}Ni_{0.21}Co_{0.08}Mn_{0.54}O_2$:** The preparation, morphology and structure characterization, and electrochemical tests of $Li_{1.17}Ni_{0.21}Co_{0.08}Mn_{0.54}O_2$ were described in a previous work.[31]

**Soft X-ray absorption spectroscopy (XAS):** Soft x-ray absorption spectroscopy was performed in the iRIXS endstation at Beamline 8.0.1 of the Advanced Light Source (ALS) at LBNL.[30] For all the samples in the 1$^{st}$ cycle, $Na_{2/3}Mg_{1/3}Mn_{2/3}O_2$ cathodes were electrochemically cycled to the representative SOC at 0.1 C rate in Swagelok cells. For the samples at $n \geq 10$ cycles, the cathodes were charged and discharged for $n$-1 cycles at 1C then were electrochemically cycled to the representative SOC at 0.1 C. The combination of different rates prevents the leakage of Swagelok cells during long-time running and ensures the accurate SOC. Electrodes were disassembled and the rinsed immediately with DMC thoroughly to lock the SOC and remove surface residue. The electrodes are then loaded into our special sample transfer chamber inside the Ar glove box. The sample transfer chamber was sealed, mounted onto the loacklock of XAS endstation for direct pump-down to avoid any air exposure effects. XAS data are collected from the side of the electrodes facing current collector in both TEY and TFY modes. All the spectra have been normalized to the beam flux measured by the upstream gold mesh. The experimental energy resolution is 0.15 eV without considering core-hole lifetime broadening.

**Mapping of Resonant Inelastic X-ray Scattering (mRIXS).** mRIXS

measurements were performed in iRIXS endstation at Beamline 8.0.1 of the ALS.[30] Data were collected through the ultra-high efficiency modular spectrometer.[63] The resolution of the excitation energy is about 0.35 eV, and the emission energy about 0.25 eV. An excitation energy step size of 0.2eV was chosen for all the maps. mRIXS were collected at each excitation energies. Final 2D images are obtained through a data process involving normalization to the beam flux and collection time, integration and combination, which has been detailed previously.[64]

**iPFY and sPFY quantifications**. Mn-$L$ iPFY spectra are obtained by formula iPFY = a/PFY$_{O\text{-}K}$ ('a' is a normalization coefficient). PFY$_{O\text{-}K}$ is obtained by integrating the Mn-$L$ mRIXS by integrating the intensity across the emission energy range from 495 eV to 510 eV with excitation energy scanning through the Mn-$L$ edges. Principles of iPFY have been explained before and briefed in this work.[30,41] Quantitative fittings of Mn-$L$ iPFY follows our established linear combination method with three standard experimental spectra[39].

The O-$K$ super-partial fluorescence yield (sPFY) signal was extracted by integrating the mRIXS intensity across the emission-energy range from 522.8 eV to 523.8 eV. This mRIXS feature has been established to be the characteristic energy range of oxygen redox activities.[20,31,33] For quantitative analysis, the spectra were scaled to the 529.7 eV peak. The area of O-$K$ sPFY 531 eV peak was integrated from 530.2 eV to 532 eV. For the cyclability analysis of extended cycles, the sPFY 531 eV peak area change is calculated by the difference between the fully charged and discharged states at the 1$^{st}$, 10$^{th}$, 50$^{th}$, and 100$^{th}$ cycles. We note that different spectral normalization does not change the conclusions in this work, because the quantified values of reversibility and cyclability only depends on the intensity contrasts (changes) of the characteristic oxygen redox features. Such a comparative intensity change could be seen directly with the sPFY raw data plots.

## ACKNOWLEDGEMENTS

The Advanced Light Source is supported by the Director, Office of Science, Office of Basic Energy Sciences, of the U.S. Department of Energy under Contract No. DE-AC02-05CH11231. We also acknowledge the support from the Fundamental Research Funds for the Central Universities of China (N110802002), the National Natural Science Foundation of China (U1504521, 51604224), Science and Technology on Reliability Physics and Application of Electronic Component Laboratory open fund (ZHD201605) and Assistant Secretary for Energy Efficiency and Renewal Energy under the Battery Materials Research (BMR) program under Contract No. DE-AC02-05CH11231. Q.L. and S.Y. acknowledge financial support from the China Scholarship Council through 111 Project No. B13029. W.Y. acknowledges the support from the Energy & Biosciences Institute through the EBI-Shell program on analyzing the Li-rich results. Z.Z. and S.S. thank the financial support of ALS doctoral fellowship.

**AUTHOR CONTRIBUTIONS**

K.D., G.L. and W.Y. conceived the project and coordinated the collaboration with contributions from all authors. K.D., J.M., G.A., C.S., Z.L. and G.L. synthesized the NMMO material and performed structural and electrochemical characterizations. K.D., J.W., Z.Z., Q.L., S.S., L.F.J.P., Y.C. and W.Y. performed spectroscopic experiments. Li-rich materials were studied with W.G. and W.C.. K.D. and W.Y. analyzed the results and prepared the manuscript with inputs from all authors.

**DECLARATION OF INTERESTS**

The authors declare no competing interests.

# Supplementary Information

# High Reversibility of Lattice Oxygen Redox Quantified by Direct Bulk Probes of both Anionic and Cationic Redox Reactions

Kehua Dai[1, 2], Jinpeng Wu[2, 3], Zengqing Zhuo[2, 4], Qinghao Li[2,5], Shawn Sallis[2, 6], Jing Mao[7], Guo Ai[8], Chihang Sun[1], Zaiyuan Li[1], William E. Gent[9], William C. Chueh[10, 11], Yi-de Chuang[2], Rong Zeng[12], Zhi-xun Shen[3], Feng Pan[4], Shishen Yan[5], Louis F. J. Piper[6], Zahid Hussain[2], Gao Liu[13, *], Wanli Yang[2, 14, *]

[1] School of Metallurgy, Northeastern University, Shenyang 110819, China

[2] Advanced Light Source, Lawrence Berkeley National Laboratory, Berkeley, California 94720, United States

[3] Geballe Laboratory for Advanced Materials, Stanford University, Stanford, California 94305, USA

[4] School of Advanced Materials, Peking University Shenzhen Graduate School, Shenzhen 518055, China

[5] School of Physics, National Key Laboratory of Crystal Materials, Shandong University, Jinan 250100, China

[6] Department of Materials Science and Engineering, Binghamton University, Binghamton, New York 13902, USA

[7] School of Materials Science and Engineering, Zhengzhou University, Zhengzhou 450001, China

[8] College of Physics and Materials Science, Tianjin Normal University, Tianjin 300387, China

[9] Department of Chemistry, Stanford University, Stanford, California 94305, USA

[10] Department of Materials Science & Engineering, Stanford University, Stanford, California 94305, USA.

[11] Stanford Institute for Materials and Energy Sciences, SLAC National Accelerator Laboratory, Menlo Park, California 94025, USA.

[12] Department of Electrical Engineering, Tsinghua University, Beijing 100084, China

[13] Energy Storage and Distributed Resources Division, Energy Technologies Area, Lawrence Berkeley National Laboratory, Berkeley, California 94720, United States

[14] Lead Contact

* Correspondence: wlyang@lbl.gov (W.Y.), gliu@lbl.gov (G.L.)

**Contents:**

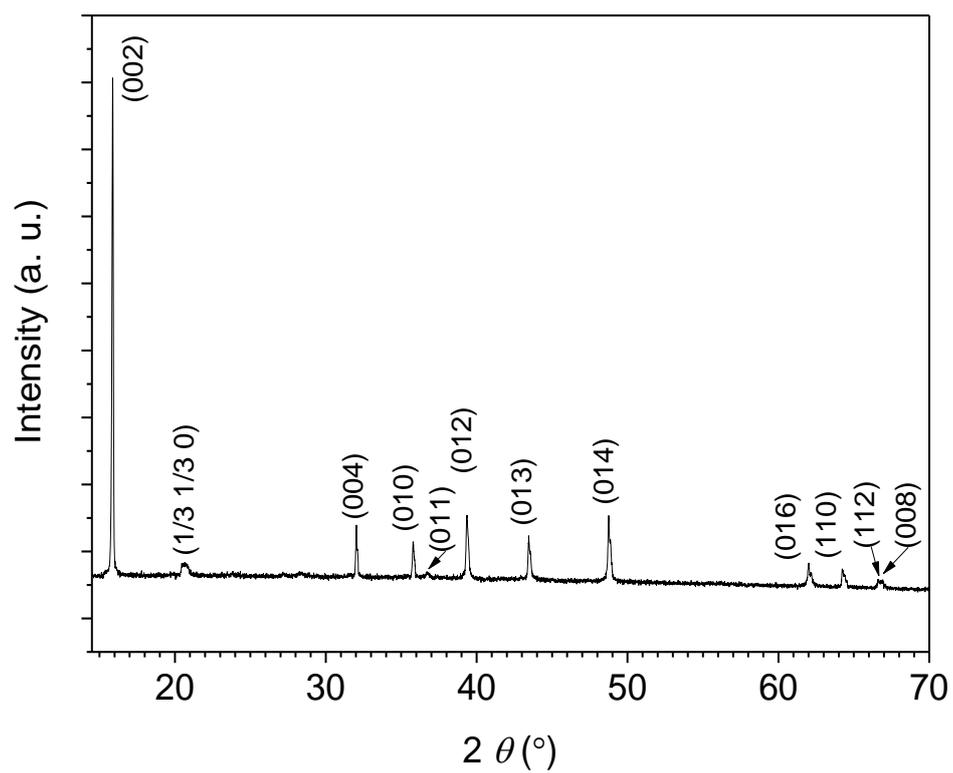

Fig. S1. The XRD pattern of the Na$_{2/3}$Mg$_{1/3}$Mn$_{2/3}$O$_2$. Scan mode: $\theta/2\theta$. Step size: 0.01°.

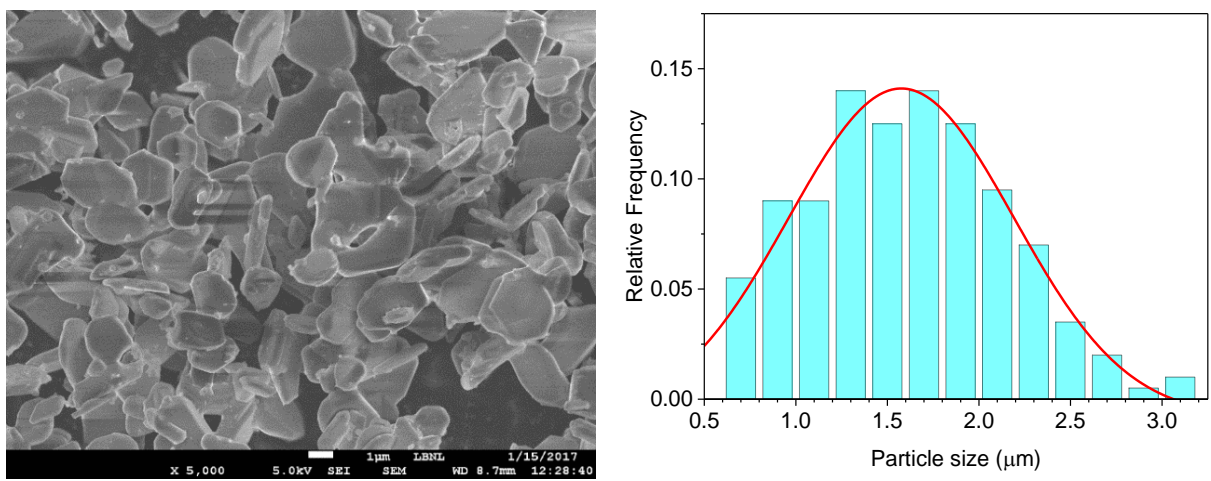

Fig. S2. The SEM image (left) and particle size distribution (right) of the $Na_{2/3}Mg_{1/3}Mn_{2/3}O_2$. The particle size distribution is counted from 200 particles in lower magnification SEM images (not shown here).

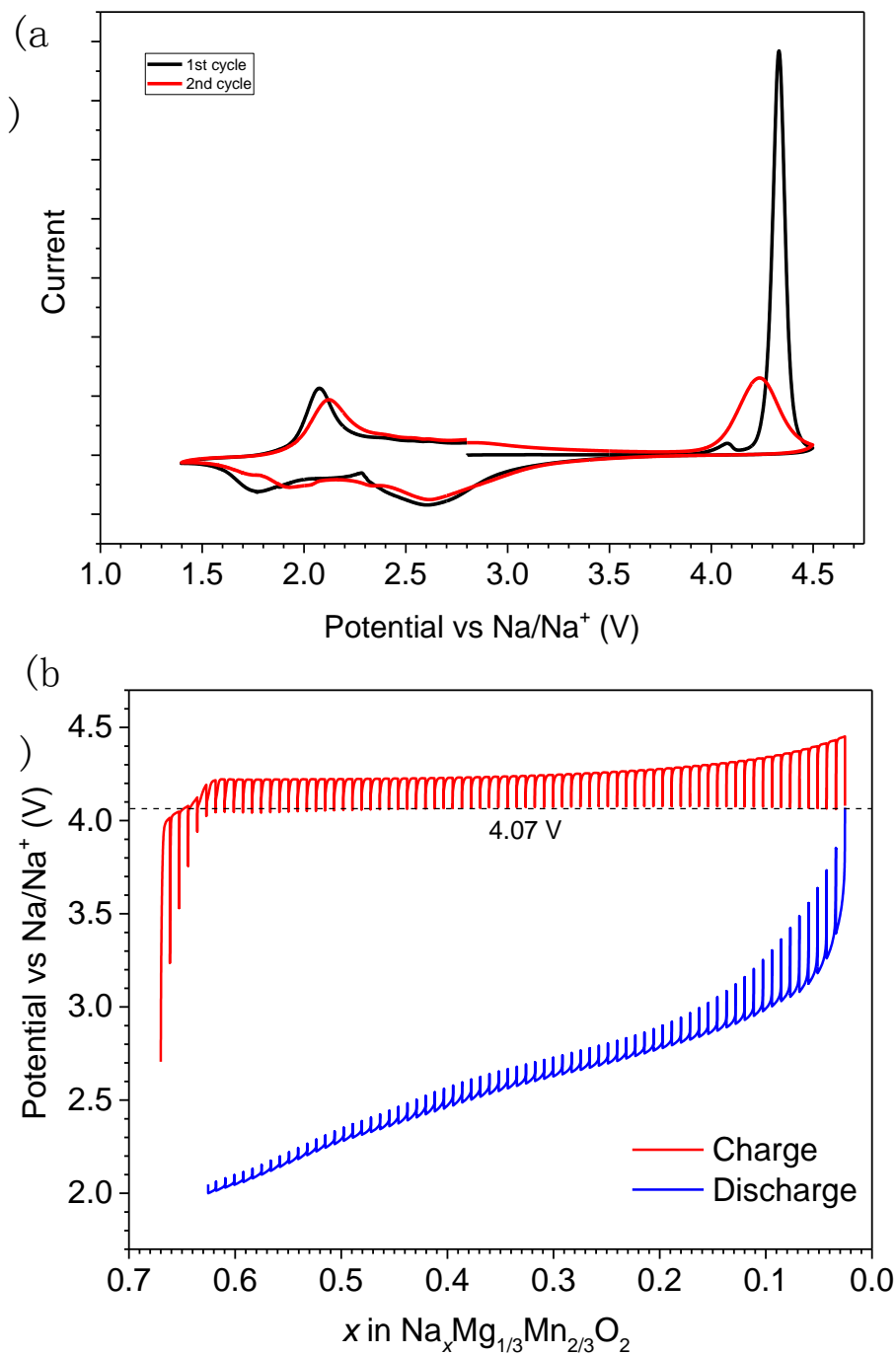

Fig. S3. (a) The cyclic voltammetry curves for 1$^{st}$ and 2$^{nd}$ cycles. Scan range: 1.4-4.5 V. Scan rate: 0.1 mV/s. (b) The galvanostatic intermittent titration technique (GITT) profiles for the 1st cycle. The charging rate is 0.1 C, each charging step is 10 min long or until cut-off voltage, and each relax step is 40 min long or until |d$V$/d$t$| < 1 mV/min. The OCV of the long charging plateau is about 4.07 V. The GITT result suggests the large voltage hysteresis is mainly due to thermodynamics.

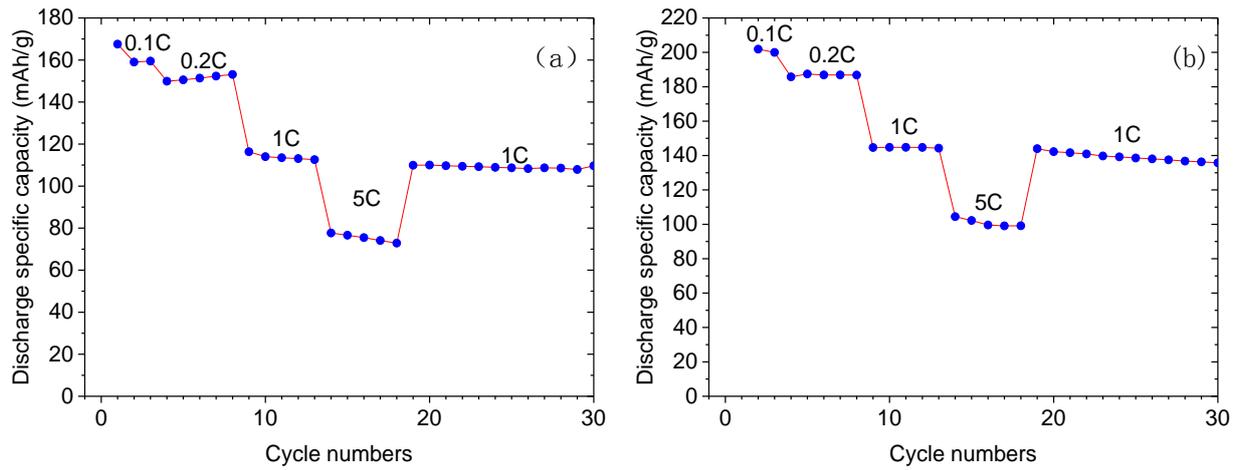

Fig. S4. The rate performance for 2.0-4.5 V (a) and 1.5-4.5 V (b). 1 C = 150 mA/g. When discharge rates are 0.1 C and 0.2 C, the charge rates are same to discharge rates. When discharge rates are 1 C and 5 C, the charge rate is 1C.

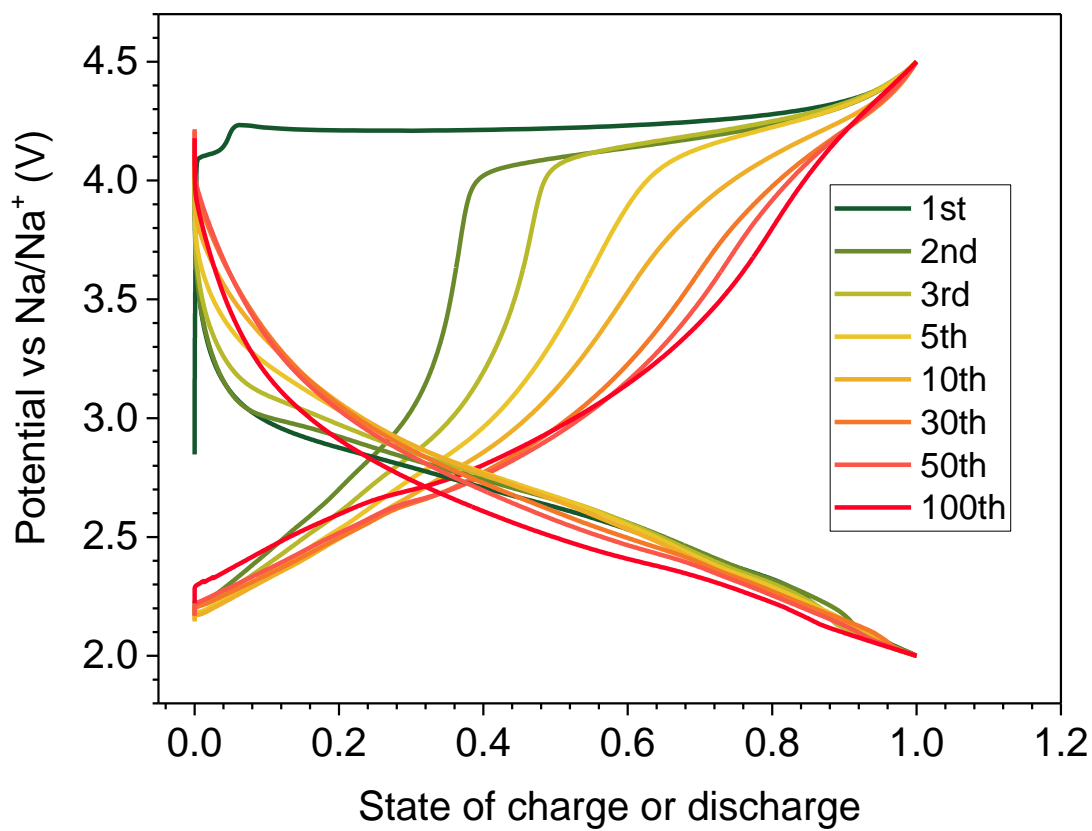

Fig. S5. The charge and discharge profiles vs. state of charge or discharge for different cycles at 0.2 C between 2.0 and 4.5 V.

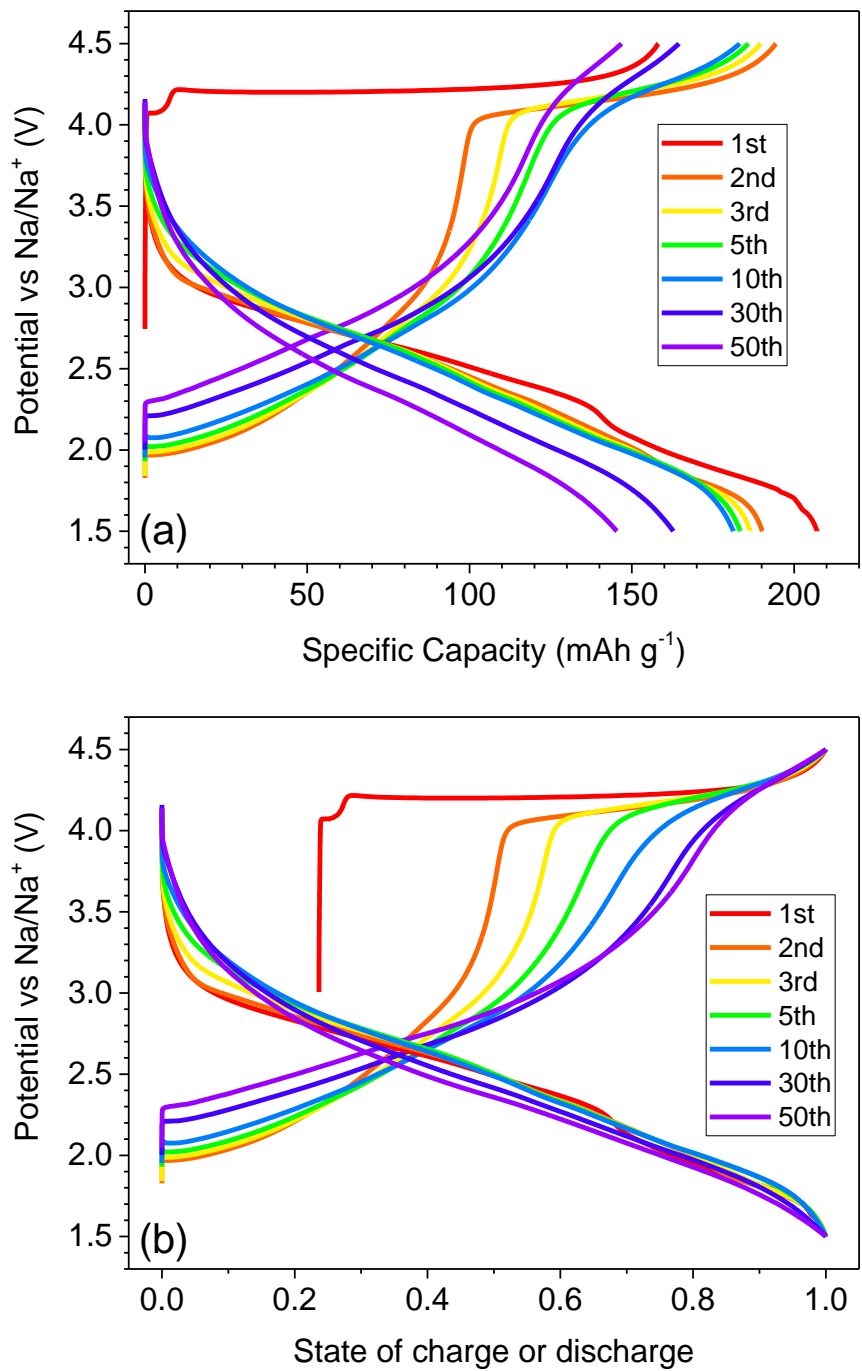

Fig. S6. The charge and discharge profiles *vs.* specific capacity (a) and state of charge or discharge (b) for different cycles at 0.2 C between 1.5 and 4.5 V.

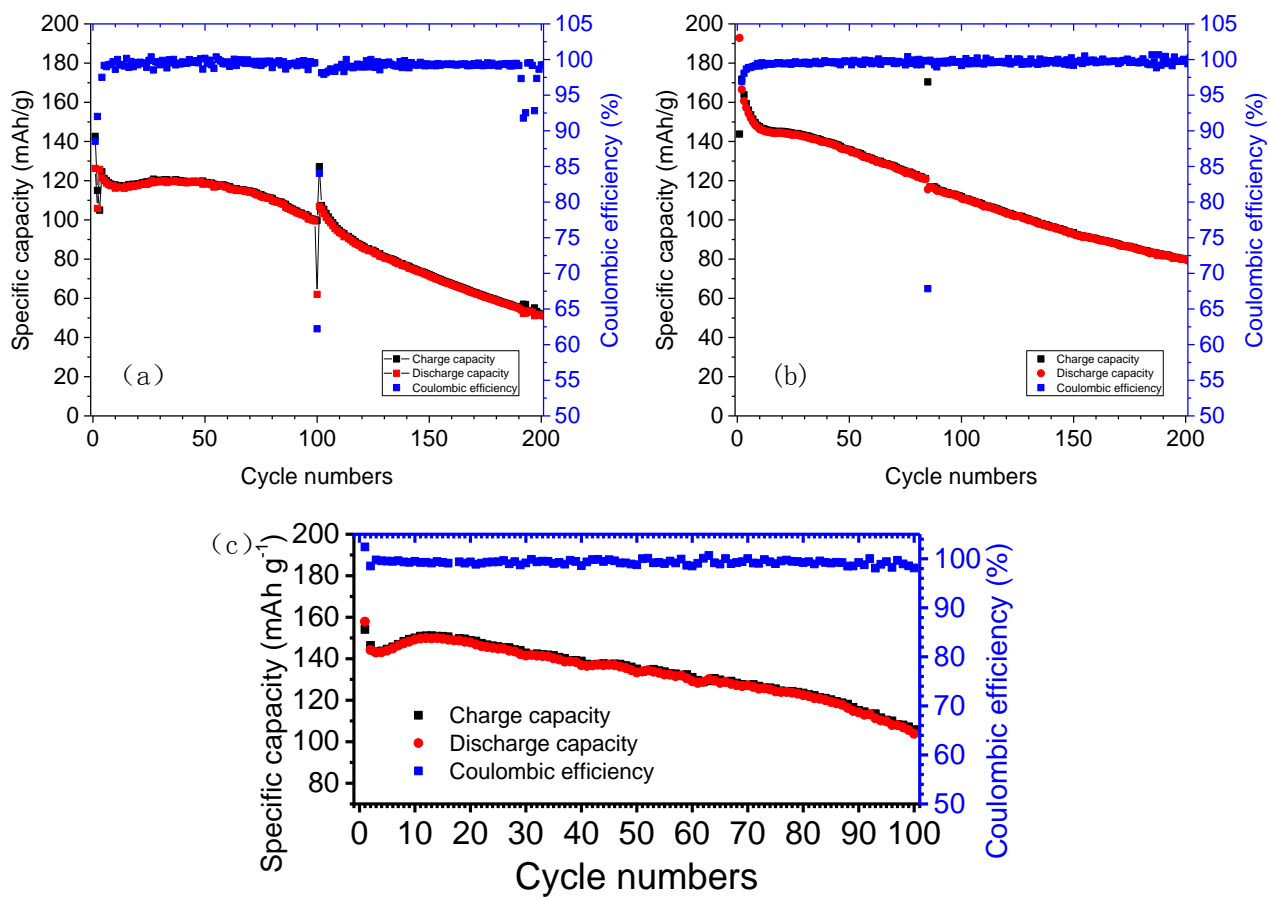

Fig. S7. The cycling performance for 2.0-4.5 V (a) and 1.5-4.5 V (b) at 1 C rate charge and discharge. The discontinuity at around 100 cycles is due to power outage of battery cycler. (c) The cycling performance for 2.0-4.5 V at 0.2 C rate charge and discharge.

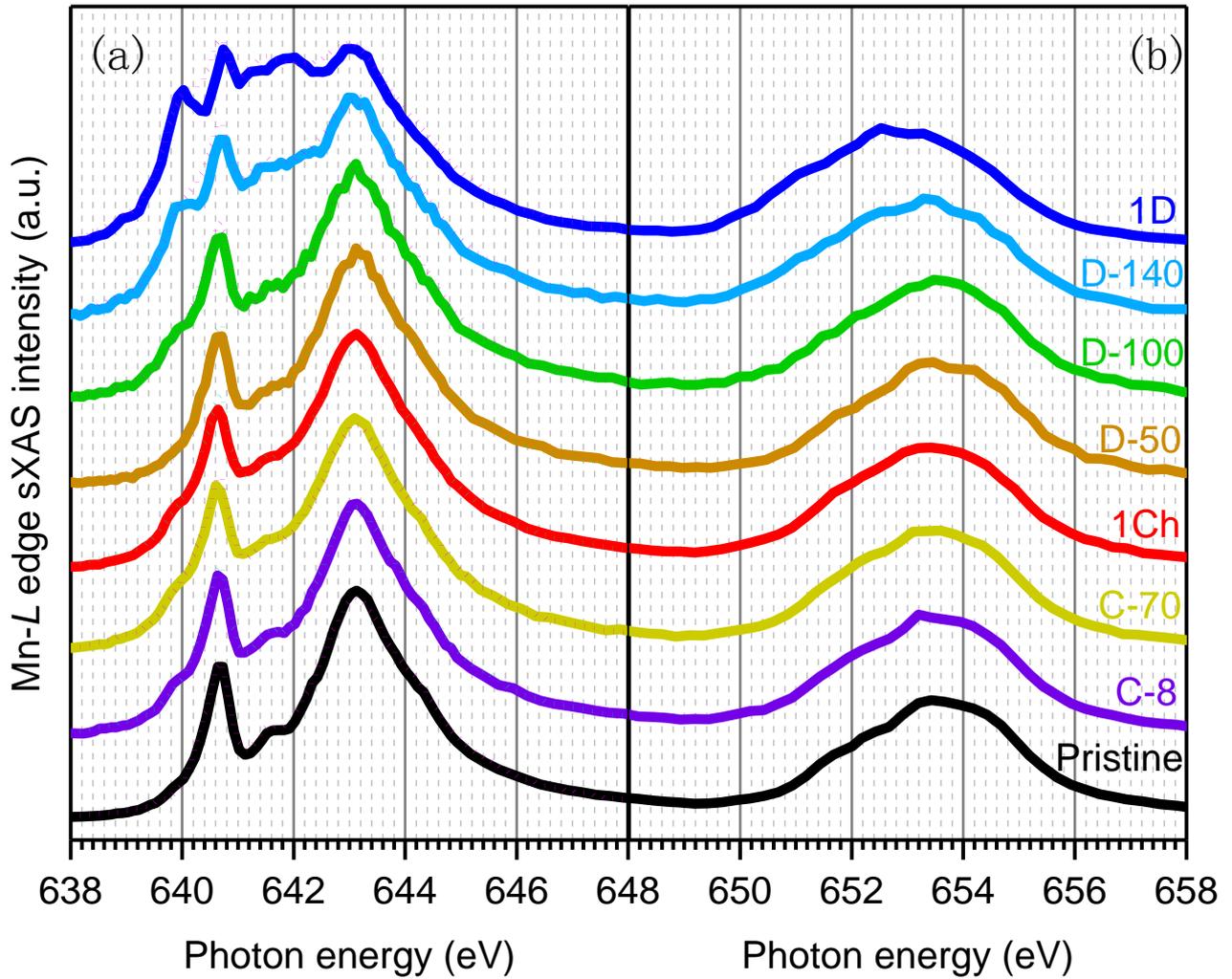

Fig. S8. (a) The Mn $L_3$-edge XAS spectra (TEY mode) collected on $Na_{2/3}Mg_{1/3}Mn_{2/3}O_2$ electrodes at different state of charge (solid lines) and the fitted curves (dotted lines). (b) The Mn $L_2$-edge XAS spectra (TEY mode) collected on $Na_{2/3}Mg_{1/3}Mn_{2/3}O_2$ electrodes at different state of charge. $L_3$ and $L_2$ edges of Mn-$L$ XAS stem from the spin-orbital splitting of the $2p^{3/2}$ and $2p^{1/2}$ states, respectively. TM $L_2$ edge suffer the intrinsic boarding from Coster-Kroniq effect[1], therefore, the sharp features in $L_3$ edge are used for quantitative fittings[2].

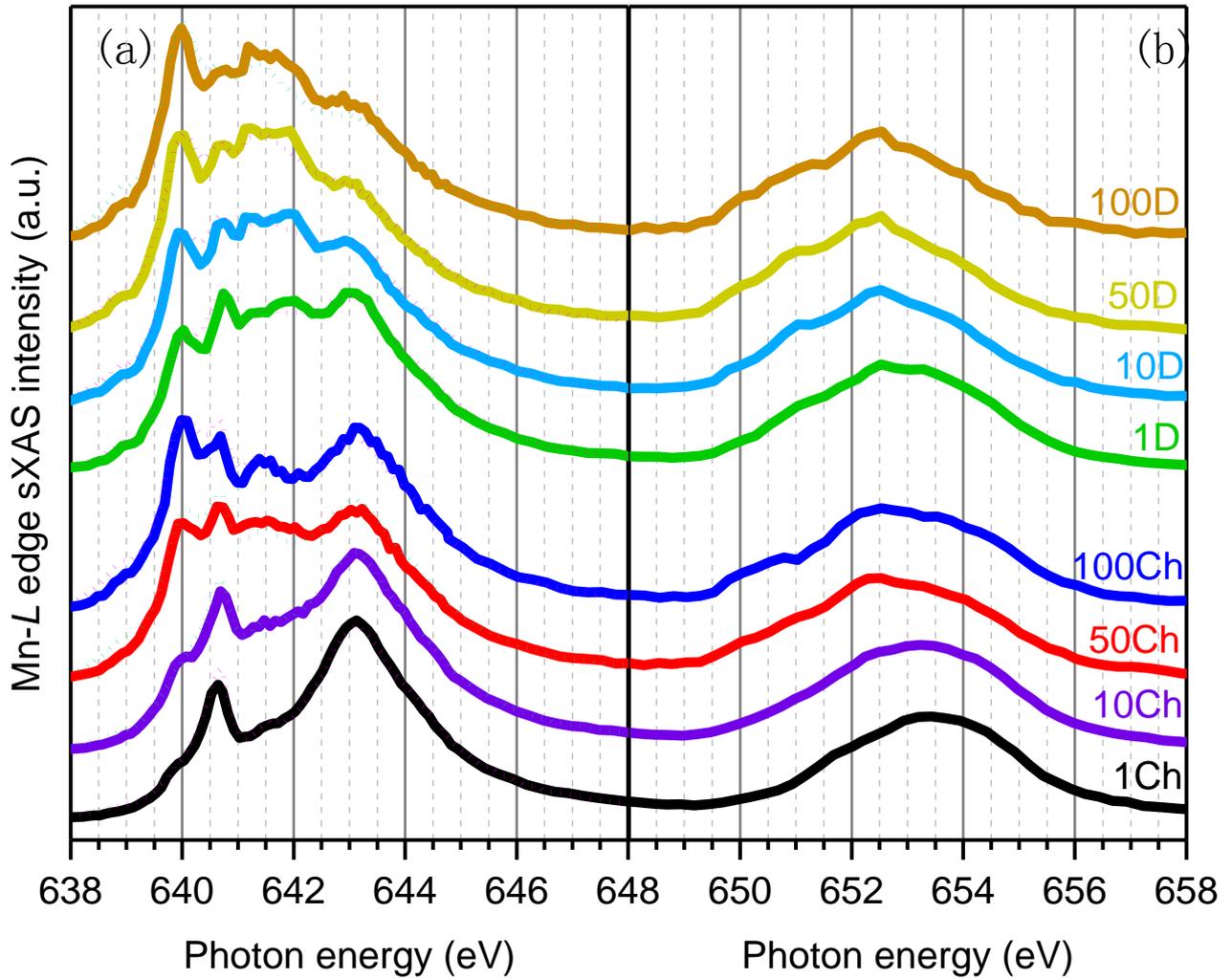

Fig. S9. (a) The Mn $L_3$-edge XAS spectra (TEY mode) collected on $Na_{2/3}Mg_{1/3}Mn_{2/3}O_2$ electrodes at different cycle (solid lines) and the fitted curves (dotted lines). (b) The Mn $L_2$-edge XAS spectra (TEY mode) collected on $Na_{2/3}Mg_{1/3}Mn_{2/3}O_2$ electrodes at different cycle. $L_3$ and $L_2$ edges of Mn-$L$ XAS stem from the spin-orbital splitting of the $2p^{3/2}$ and $2p^{1/2}$ states, respectively. TM $L_2$ edge suffer the intrinsic boarding from Coster-Kroniq effect[1], therefore, the sharp features in $L_3$ edge are used for quantitative fittings[2].

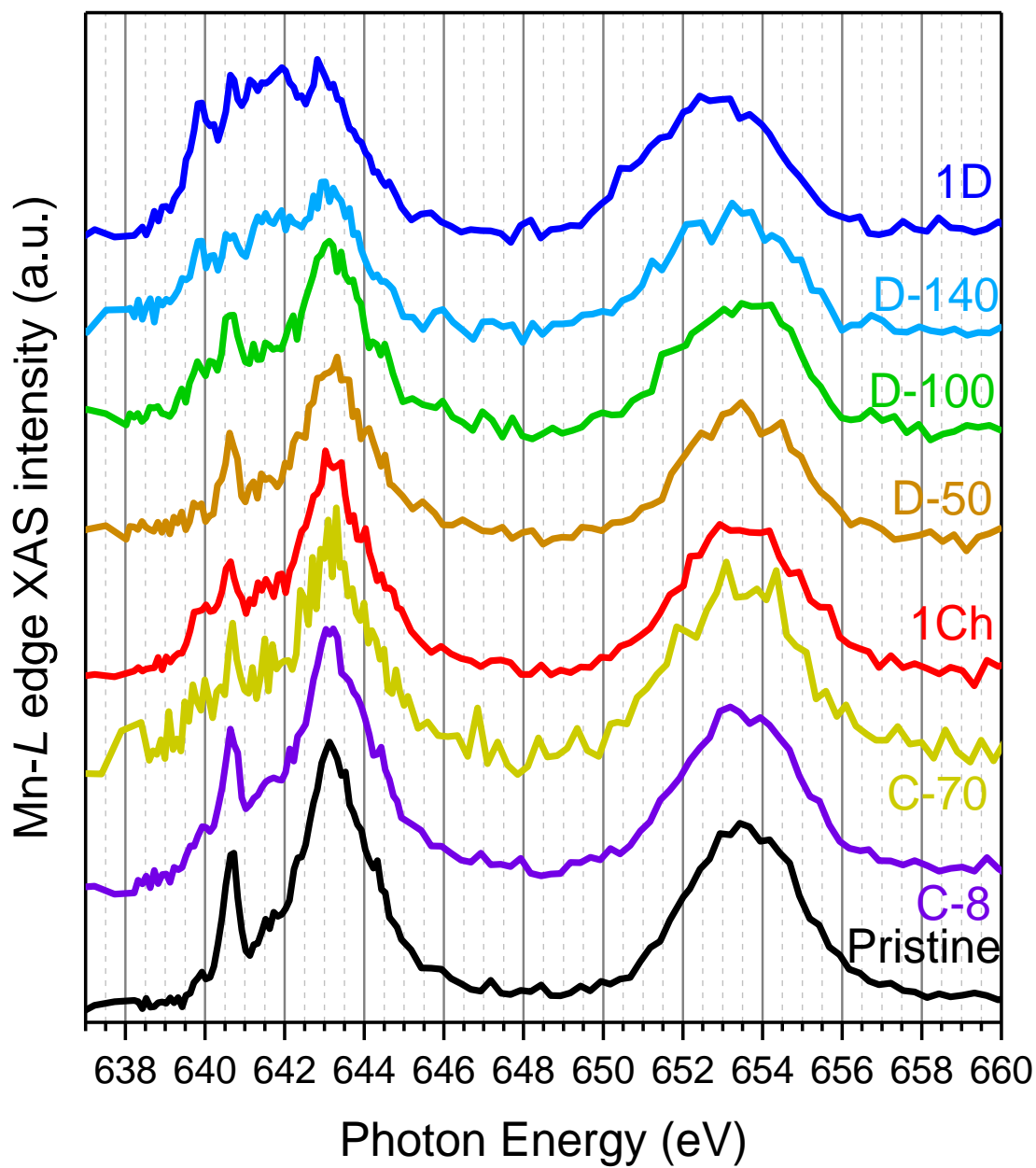

Fig. S10. The Mn XAS spectra (TFY mode) collected on $Na_{2/3}Mg_{1/3}Mn_{2/3}O_2$ electrodes at different state of charge. Distortions of the $L_3$ edge (638-648 eV) are largely due to the saturation effect because Mn edge is sitting on the oxygen absorption background [3]. However, the $L_2$ feature (648-658 eV) shift is generally consistent with the Mn valence analysis.

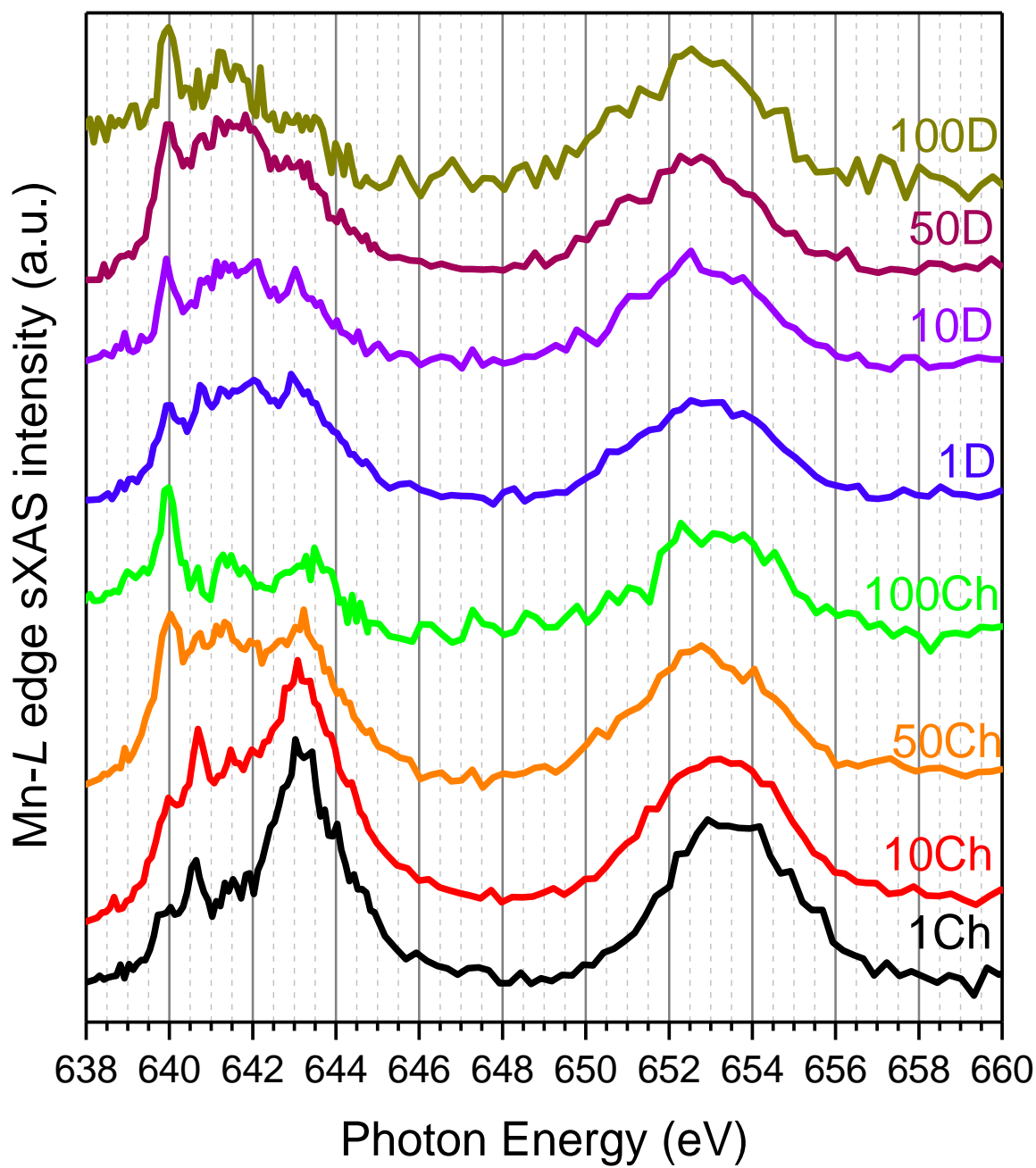

Fig. S11. The Mn XAS spectra (TFY mode) collected on $Na_{2/3}Mg_{1/3}Mn_{2/3}O_2$ electrodes at different cycle. Distortions of the $L_3$ edge (638-648 eV) are largely due to the saturation effect because Mn edge is sitting on the oxygen absorption background [3]. However, the $L_2$ feature (648-658 eV) shift is generally consistent with the Mn valence analysis.

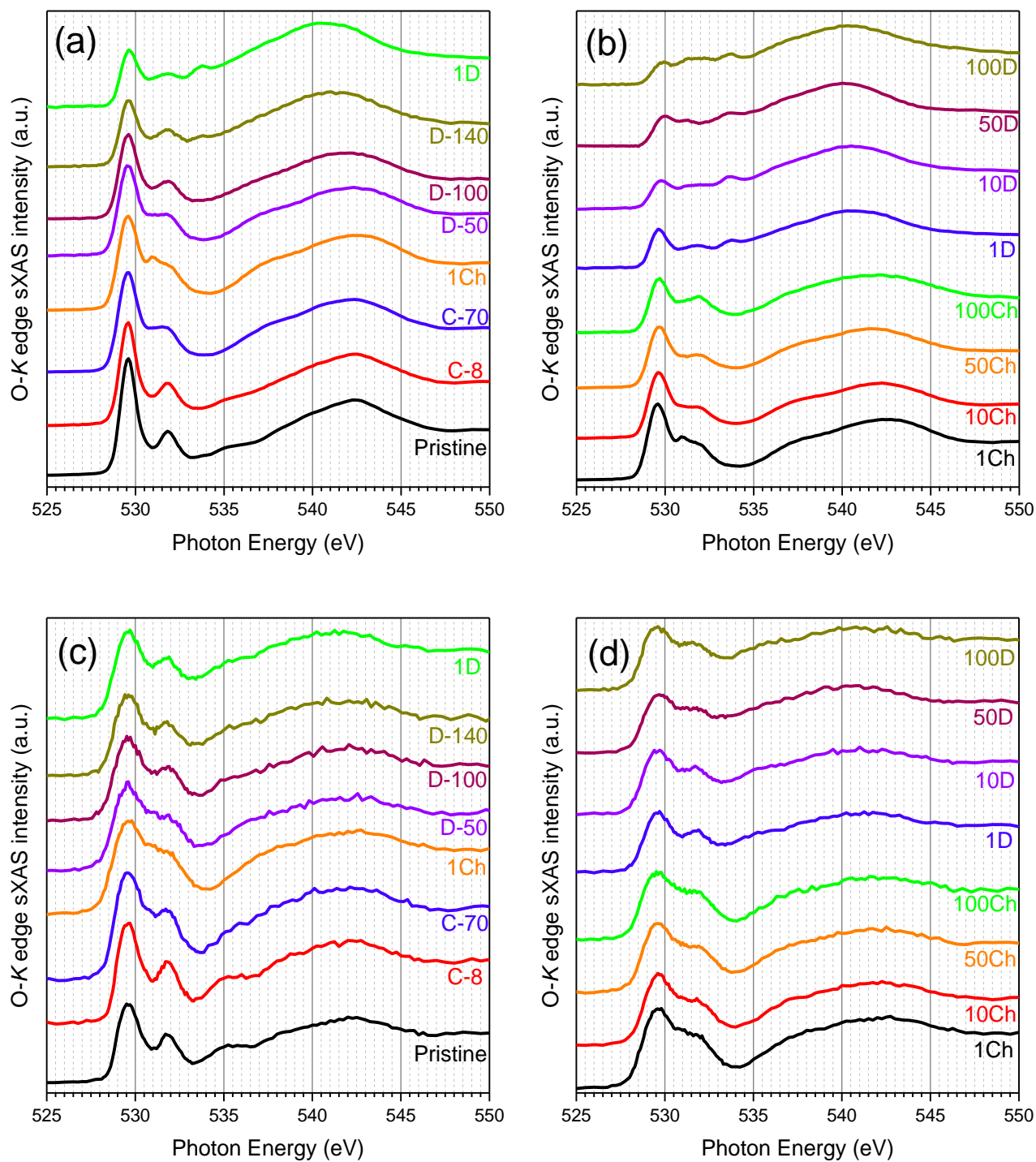

Fig. S12. The O *K*-edge sXAS spectra collected in TEY (a, b) and TFY (c, d) modes on Na$_{2/3}$Mg$_{1/3}$Mn$_{2/3}$O$_2$ electrodes at different SOC during the initial cycle (a, c), and the charged (Ch) and discharged (D) states at the 1st, 10th, 50th, 100th cycles (b, d). The features at 528-533 eV display dramatic variations upon electrochemical cycling, however, represent the change on Mn states (lineshape) and Mn-O hybridization strength (intensity). XAS results are used only as supplementary information in this work.